\documentclass[letterpaper]{article} 
\usepackage{aaai2026}  
\usepackage{times}  
\usepackage{helvet}  
\usepackage{courier}  
\usepackage[hyphens]{url}  
\usepackage{graphicx} 
\usepackage{natbib}  
\usepackage{caption} 
\usepackage{algorithm}
\usepackage{algorithmic}
\usepackage{amsmath}
\usepackage{amsfonts}
\usepackage{newfloat}
\usepackage{listings}

\usepackage{tabularx}
\newcolumntype{Y}{>{\raggedright\arraybackslash}X} 

\usepackage{multirow}
\usepackage{rotating}
\usepackage{bm}
\usepackage{booktabs}

\newtheorem{definition}{Definition}
\newtheorem{lemma}{Lemma}
\newtheorem{proof}{Proof}
\usepackage{easyReview}
\DeclareCaptionStyle{ruled}{labelfont=normalfont,labelsep=colon,strut=off} 
\floatstyle{ruled}
\newfloat{listing}{tb}{lst}{}
\floatname{listing}{Listing}

\usepackage[table]{xcolor}
\usepackage{array}

\newcolumntype{g}{>{\columncolor[gray]{0.9}}r} 

\title{Towards Information-Optimized Multi-Agent Path Finding: A Hybrid Framework with Reduced Inter-Agent Information Sharing}
\author{
    Bharath Muppasani,
    Ritirupa Dey, Risha Patel,
    Biplav Srivastava,
    Vignesh Narayanan
}
\affiliations{
    University of South Carolina, USA\\
    bharath@email.sc.edu, deyr@email.sc.edu, biplav.s@sc.edu, vignar@sc.edu
}

\usepackage{bibentry}

\begin{document}

\maketitle

\begin{abstract}

    Multi-agent pathfinding (MAPF) remains a critical problem in robotics and autonomous systems, where agents must navigate shared spaces efficiently while avoiding conflicts. Traditional centralized algorithms with global information provide high-quality solutions but scale poorly in large-scale scenarios due to the combinatorial explosion of conflicts. Conversely, distributed approaches that have local information, particularly learning-based methods, offer better scalability by operating with relaxed information availability, yet often at the cost of solution quality. In realistic deployments, information is a constrained resource: broadcasting full agent states and goals can raise privacy concerns, strain limited bandwidth, and require extra sensing and communication hardware, increasing cost and energy use. We focus on the core question of how MAPF can be solved with \emph{minimal} inter-agent information sharing while preserving solution feasibility.
    To this end, we present an information-centric formulation of the MAPF problem and introduce a hybrid framework, IO-MAPF, that integrates decentralized path planning with a lightweight centralized coordinator. In this framework, agents use reinforcement learning (RL) to plan independently, while the central coordinator provides minimal, targeted signals, such as static conflict-cell indicators or short conflict trajectories, that are dynamically shared to support efficient conflict resolution.
    We introduce an Information Units (IU) metric to quantify information use and show that our alert-driven design achieves \bm{$2\times$} to \bm{$23\times$} reduction in information sharing, compared to the state-of-the-art algorithms, while maintaining high success rates, demonstrating that reliable MAPF is achievable under strongly information-restricted, privacy-preserving conditions. We demonstrate the effectiveness of our algorithm using both simulation and hardware experiments.


\end{abstract}


\section{Introduction \label{main: intro}}

Multi-Agent Path Finding (MAPF) addresses the fundamental problem of computing collision-free trajectories for multiple agents navigating a shared environment. Solving it effectively is crucial for a wide range of real-world applications, ranging from the deployment of automated robotic swarms in warehouses to performing autonomous vehicle coordination.  
Despite its practical significance, solving MAPF is computationally demanding, classified as NP-hard in its general form, which can render traditional centralized search methods intractable as the number of agents or the complexity of the environment increases \cite{mapf-empiricalhardness-bluesky,sartoretti2019primal}.

\begin{figure}[!t]
    \centering
    \includegraphics[width=1\linewidth]{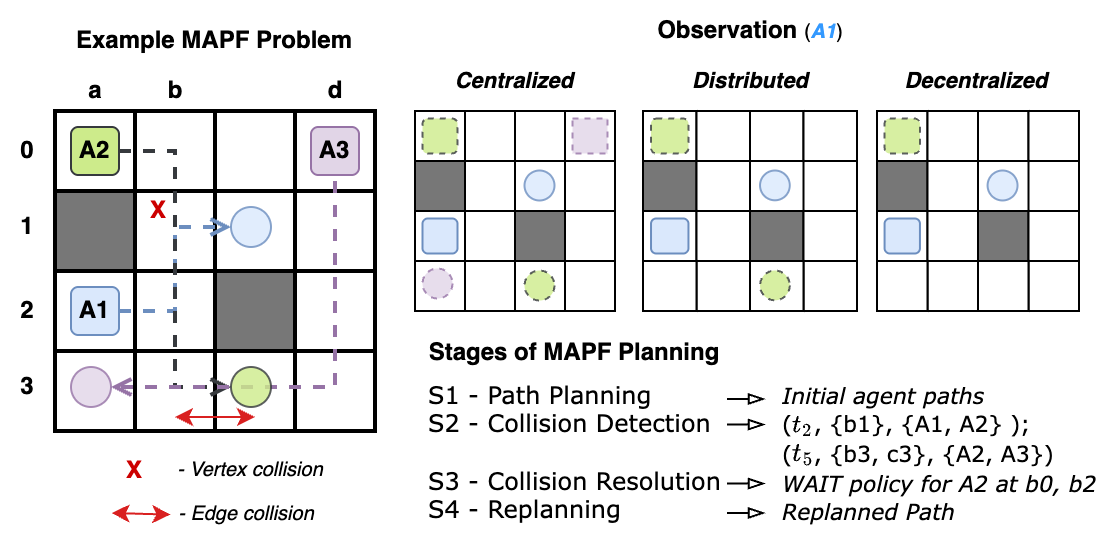}
    \caption{\footnotesize An example MAPF problem and our four-stage planning pipeline. 
    \textbf{Left:} Three agents (A1, A2, A3) navigate a grid with static obstacles (dark gray). The diagram illustrates a future vertex collision (red X), where two agents would occupy the same cell, and an edge collision (red arrows), where agents would swap adjacent cells. 
    \textbf{Top Right:} The varying levels of information available to Agent A1 under centralized (all agent positions and goals), distributed (nearby agent positions and goals), and decentralized (only nearby agent positions) paradigms. 
    \textbf{Bottom Right:} The four stages of our framework, from initial Path Planning (S1) to Collision Detection (S2), Resolution (S3), and Replanning (S4).}
    \label{fig:mapf}
    \vspace{-0.5em}
\end{figure}

The MAPF literature has explored various coordination paradigms, each with different implications for information availability and system performance. Centralized approaches, such as Conflict-Based Search (CBS)~\cite{Sharon2015} and its efficiency-focused variant ICBS~\cite{boyarski2015icbs}, typically assume full 
knowledge of the environment and agent states and compute collision-free paths based on this global information available. 
However, as the number of agents or the size of the environment increases, joint planning and conflict resolution become computationally expensive. Additionally, requiring agents to share full information raises privacy concerns, limiting applicability in confidential settings and imposing additional compute requirements in real-world deployments, motivating strategies that reduce reliance on global information. Early decoupled methods like $M^*$~\cite{WagnerChoset2011} begin with individual plans and only coordinate agents in conflict, offering scalability but often sacrificing optimality or completeness. More recently, Multi-Agent Reinforcement Learning (MARL) has emerged to address coordination and partial observability in dynamic settings. 
For example, PRIMAL~\cite{sartoretti2019primal} trains agents to plan using partial views, learning implicit coordination, while methods like FOLLOWER~\cite{Skrynnik2024} reduce explicit communication by relying on global heuristic maps. These decentralized methods improve scalability but may reduce solution quality and leave conflicts unresolved due to limited information.

To address these limitations, we first formulate the MAPF problem from an information-centric perspective and with the aim of reducing information exchange under this setting, we propose \emph{Information-Centric MAPF (IC--MAPF)}, a novel hybrid MAPF framework that combines decentralized planning with a lightweight centralized coordinator, 
that \emph{minimizes} inter-agent information sharing while maintaining solution feasibility. Our findings show that minimal, targeted alerts are sufficient, reducing the total information load by an estimated \bm{$2\times$} to \bm{$23\times$}
compared to the state-of-the-art communication efficient algorithms (see Table~\ref{tab:search-vs-learning}). In our approach, agents primarily rely on decentralized, reinforcement learning-based neural network planners that operate on local observations (e.g., position coordinates), eliminating the need for global information as well as local egocentric maps. A central coordinator oversees agent trajectories, intervening selectively by dynamically sharing targeted information to prompt localized re-planning when conflicts are anticipated. This work's contributions are centered on this selective coordination strategy. We introduce (1) \textbf{a information-centric MAPF formulation} and identify the information utilized by various MAPF strategies, \textbf{(2) IC--MAPF - a novel hybrid framework} with an on-demand alert mechanism that substantially reduces the amount of global information an agent must access for successful planning, and \textbf{(3) a new metric, Information Units (IU)}, to quantify the information available to each agent during planning.


Our evaluation is guided by two key research questions: \textbf{\emph{(RQ1)}} Given the observation constraints of a decentralized setup, can an effective MAPF algorithm be created with one agent knowing nothing about other agents? If not, what information must it need at a minimum? and \textbf{\emph{(RQ2)}} How does the proposed hybrid method compare to leading alternative search- and learning-based approaches in terms of performance, solution quality, and scalability?

\section{Background and Literature Review \label{main: lit}}

\subsection{Multi‑Agent Path Finding}
Let \(G = (V, E)\) be an undirected graph, where \(V\) is the set of vertices (grid cells) and \(E\subseteq V\times V\) is the set of edges connecting adjacent cells.  A team of \(n\) agents \(A=\{a_1,\dots,a_n\}\) must move from start vertices \(s_i\in V\) to goal vertices \(g_i\in V\), where $(s_i,g_i) \ne (s_j,g_j) ,\forall i \ne j: i,j \in \{1,...,n\}$.  Time is discretized into steps \(t=0,1,2,\dots\), and at each step, an agent may either move along an edge or wait.  A path for agent \(a_i\) is a sequence \(\pi_i=(v^i_0,v^i_1,\dots,v^i_{T_i})\) with \(v^i_0=s_i\) and \(v^i_{T_i}=g_i\).  A solution to MAPF is \(\Pi=\{\pi_1,\dots,\pi_n\}\), and it is collision‑free if for all \(i\neq j\) and all \(t\),
\(
v^i_t \neq v^j_t\) (vertex collision free) and 
\(
(v^i_t,v^i_{t+1}) \neq (v^j_{t+1},v^j_t)\)
(edge collision free). The primary goal in standard MAPF is typically to find a path set \(\Pi\) that is collision-free \cite{Sharon2015}. Common efficiency objectives include minimizing the makespan, \(\max_i T_i\), minimizing the sum of individual completion times (sum-of-costs), \(\sum_i T_i\).


An important modeling choice in MAPF concerns how agents are treated after reaching their goals at potentially different times. Two common settings are used \cite{stern2019multi}. In the \emph{stay-at-target} setting, an agent remains at its goal vertex after arrival and continues to occupy that location until all agents have finished, potentially creating conflicts for others. In contrast, the \emph{disappear-at-target} setting removes an agent from the environment immediately upon goal arrival, eliminating any further interactions. In this work, we adopt the \emph{stay-at-target} setting, which is appropriate for persistent agents operating in shared environments.

\subsection{Coordination Paradigms}

As formalized by Sharon et al.\ \cite{Sharon2015}, solution approaches to MAPF problems can be categorized based on their coordination strategy. We extend this by focusing on three key elements: state observability (global vs.\ local), communication (allowed vs.\ none), and control (centralized vs.\ decentralized).
In \emph{centralized} paradigms, a global planner with full observability of the entire state controls all agents, and communication is implicit through this central controller.
By contrast, other approaches grant agents local control over their actions. Within this category, we draw a key distinction based on communication:
In \emph{distributed} approaches, agents that plan their own paths are allowed to explicitly exchange information, such as local observations, intended paths, or goals, with other agents to achieve cooperative behavior.
{\bf In \emph{decentralized} MAPF, as defined in this work, inter-agent communication during execution is eliminated}. Agents plan and act entirely independently, relying only on their own path information. Finally, \emph{hybrid} frameworks, like the one we propose, combine these elements. They typically employ a central coordination mechanism that has full observability to detect conflicts and selectively intervene, while agents otherwise plan in a decentralized manner.
In our work, we adopt a hybrid framework with a customized information sharing mechanism as described in Section \ref{main: method}.
\subsection{Literature Review}
\textbf{Search-based Methods:} Classical MAPF research progressed from centralized optimal solvers to more scalable search frameworks. Early methods such as $M^*$ dynamically couple agents only when conflicts arise \cite{WagnerChoset2011}, while ICTS allocates costs to agents and searches for conflict-free combinations of single-agent paths, reducing joint-space complexity \cite{Sharon2013}. A major milestone was CBS, which plans independently for each agent using $A^*$ and resolves conflicts by branching on constraints, guaranteeing optimality \cite{Sharon2015}. Its variants, including ICBS, further accelerate search through cardinality heuristics and meta-agent merging \cite{boyarski2015icbs}. For large-scale problems, suboptimal approaches like LNS generate an initial solution (often via prioritized planning) and iteratively repair conflicting subsets \cite{LiAAAI22}. PIBT \cite{okumura2022priority} introduces adaptive prioritization for local agent movements, and LaCAM \cite{okumura2023lacam} combines low-level constraint extraction with high-level location-sequence search. Despite their efficiency gains, these methods still face scalability limits with growing agent counts and rely heavily on global information.

\textbf{Learning-based Methods:} Learning-based MAPF methods focus on handling partial observability and limited communication through decentralized policies trained under the centralized training and decentralized execution (CTDE) paradigm. PRIMAL and PRIMAL2 combine imitation learning from an expert centralized solver with reinforcement learning (PRIMAL2 further improving local observations), enabling scalable decentralized execution \cite{sartoretti2019primal, Damani2021PRIMAL2}. SCRIMP employs a transformer-based communication module that supports coordination among agents with restricted field-of-view (FOV), allowing decentralized conflict avoidance without a central coordinator \cite{Wang2023SCRIMP}. SYLPH \cite{he2025social} enables agents to infer their Social Value Orientation (SVO), a measure of selfishness or altruism, based on situational context, and adapt their behavior toward the most influential agent in the system. 
ALPHA \cite{he2024alpha} blends accurate proximal sensing with fuzzy global cues to guide agents beyond purely local reasoning, mitigating local myopia. SIGMA \cite{liao2025sigma} uses sheaf-theoretic local consensus to capture geometric cross-dependencies among agents. 
In real deployments, each agent must carry onboard sensors (e.g., RGB-D or LiDAR) and maintain communication channels for goal disclosure, introducing extra computation and potential privacy risks. While adaptable and fully decentralized, learning-based methods often require extensive training, may generalize poorly, and provide weaker guarantees on solution quality or completeness when relying solely on local information.


\textbf{Hybrid Methods:} Hybrid planning–learning approaches aim to combine the strengths of both paradigms. FOLLOWER \cite{Skrynnik2024} uses congestion-aware $A^*$ planning to generate subgoals, followed by decentralized RL for replanning. DCC \cite{ma2021learning} and EPH \cite{tang2024ensembling} augment local FOV sensing with selective inter-agent communication, while DHC \cite{ma2021distributed} exchanges information with $K$ nearest neighbors to improve coordination. EPH further employs ensemble-based inference and switches to fast $A^*$ when no agents appear in FOV. More recent systems such as LNS2+RL \cite{WangAAAI25} combine LNS with MARL for early conflict resolution before refining solutions via search. Although effective, these hybrid designs often require complex integrations and specific information dependencies, and typically trade optimality for scalability.

To enable a structured comparison across search-based, learning-based, and hybrid MAPF methods, we adopt a four-stage pipeline: S1 (Agent Planning), S2 (Collision Detection), S3 (Collision Avoidance Policy), and S4 (Agent Replanning). Within this view, CBS performs decentralized $A^*$ for S1/S4 and uses a centralized high-level solver for S2/S3, ensuring optimality. LNS centralizes all four stages, path generation, conflict detection, repair, and subset replanning, trading optimality for scalability. Learning-based approaches rely on local observations to decentralize S1–S4, though real deployments typically require onboard sensing (e.g., cameras, LiDAR) and inter-agent communication, increasing cost and computation. Hybrid methods occupy the middle ground: FOLLOWER uses global congestion maps with $A^*$ for S1 and decentralized RL for S2–S4, while LNS2+RL combines centralized LNS for S1–S3 with a hybrid RL/prioritized-planning strategy for S4. A detailed categorization is provided in Table~\ref{tab:lit_comparison} (Appendix Section~\ref{ap: lit}).

Overall, existing MAPF paradigms face key trade-offs: centralized methods struggle with scalability and privacy; purely learning-based methods can weaken solution quality; and current hybrids still rely on substantial centralized coordination or costly sensing. {\em The hybrid framework, IC--MAPF, introduced in Section~\ref{main: method} is designed to overcome these limitations} by reducing inter-agent information sharing while preserving feasibility, thereby avoiding the communication and sensing overhead typical of learning-based systems.

\vspace{-0.1in}
\section{Methodology \label{main: method}}


\subsection{Information-Centric MAPF}
\begin{table*}[t] 
\centering
\renewcommand{\arraystretch}{1.1}
\resizebox{\textwidth}{!}{%
\begin{tabular}{l l p{13.5cm}}
\toprule
\textbf{Algorithm} & \textbf{Category} & \textbf{Information Used by Agent \(i\)} \\
\midrule
CBS/ EECBS/ LaCAM/ LaCAM*/ LNS/ LNS2 
& Centralized (Search-Based)
& \( \mathcal{I}_i^{\text{central}}(t) = (\mathcal{M},\, \mathcal{S}_i = (s_i^t, g_i^t, \pi_{i|t_0:t_{\max}}),\, \mathcal{O}_j = (s_j^t, g_j^t ,\pi_{j|t_0:t_{\max}})_{j \neq i} ) \) \\
\midrule
MRCDRL / AMPPO-PP
& Centralized (Learning-Based)
& \( \mathcal{I}_i^{\text{central}}(t) = (\mathcal{M},\, \mathcal{S}_i = (s_i^t, g_i^t, \pi_{i|t_0:t_{\max}}),\, \mathcal{O}_j = (s_j^t, g_j^t ,\pi_{j|t_0:t_{\max}})_{j \neq i} ) \) \\
\midrule
PIBT / CS-PIBT /
& Distributed (Search-Based)
& \( \mathcal{I}_i^{\text{distributed}}(t) = (\mathcal{M}_i^k,\, \mathcal{S}_i = (s_i^t, g_i^t, \pi_{i|t_s:t_r}),\, \mathcal{O}_j = \{ (s_j^t, g_j^t, \pi_{j|t_s:t_r}) \}_{j \in \mathcal{N}_i(t)} ) \) \\
\midrule
SACHA / PRIMAL / SCRIMP
& Distributed (Learning-Based)
& \( \mathcal{I}_i^{\text{distributed}}(t) = (\mathcal{M}_i^k,\, \mathcal{S}_i = (s_i^t, g_i^t, \pi_{i|t_s:t_r}),\, \mathcal{O}_j = \{ (s_j^t, g_j^t, \pi_{j|t_s:t_r}) \}_{j \in \mathcal{N}_i(t)} ) \) \\
\midrule
FOLLOWER
& Decentralized (Learning-Based)
& \( \mathcal{I}_i^{\text{decentral}}(t) = (\mathcal{M}_i^k,\, \mathcal{S}_i = (s_i^t, g_i^t, \pi_{i|t_s:t_r})) \) \\
\midrule
MA-CBS/ M*/ HiPP-MAPP/ DCC/ DHC/ EPH/ \textbf{IC--MAPF} 
& Hybrid 
& \( \mathcal{I}_i^{\text{hybrid}}(t) = (\mathcal{M}_i,\, \mathcal{S}_i = (s_i^t, g_i^t, \pi_{i|t_s:t_r}),\, \mathcal{O}_j = \{ (s_j^t, g_j^t, \pi_{j|t_s:t_r}) \}_{j \in \mathcal{N}_i(t)} , \mathcal{E}_{ij}) \) \\
\bottomrule
\end{tabular}%
}
\caption{Categorization of MAPF algorithms by information availability, based on the unified model \(\mathcal{I}_i(t) = (\mathcal{M}_i, \mathcal{S}_i, \mathcal{O}_j)\).}
\label{tab:information_mapf}
\end{table*}


{In an MAPF problem, each agent 
\(i \in \{1,\dots,N\}\) relies on a set of information 
components to safely navigate from its start to its goal. 
This includes \emph{environmental information}, represented 
by the entire map or a subset of the map \(\mathcal{M}_i = (V, E, O)\) of vertices, edges, 
and static obstacles; \emph{self information}, consisting of 
the agent’s current state \(s_i^t \in V\), goal \(g_i^t \in V\), 
and its augmented policy over a time interval $\{t_s,...,t_r\}$ given by $\pi_{i|t_s:t_r}$, which can be compactly denoted by $\mathcal{S}_i =(s_i^t,g_i^t,\pi_{i|t_s:t_r})$; 
and some amount of 
\emph{other-agent information}, specifically their current positions, goal positions 
and the immediate intended actions following their individual policies, 
$\mathcal{O}_j=(s_j^t,g_j^t, \pi_{j|t_s:t_r} )_{j \in \mathcal{N}_i(t)}$, where $\mathcal{N}_i(t)$ denotes the set of agents in the neighborhood of agent $i$, required to 
avoid vertex and edge collisions. Collectively, the 
information accessible to agent \(i\) at any timestep t, is written as
$
\mathcal{I}_i(t)
=
\bigl(
\mathcal{M}_i,\, 
\mathcal{S}_i,
\mathcal{O}_j
\bigr).
$
While this set captures the general information required for 
MAPF, different algorithmic paradigms make use of 
\(\mathcal{I}_i(t)\) in distinct ways. In \emph{centralized} 
algorithms, a global solver gathers the full map information and the {full joint 
information } \(\{ s_i^t,\, g_i^t \}_{i=1}^N\) and computes 
conflict-free paths \(\{ \pi_i \}_{i=1}^N\); during execution, 
each agent requires only
$\mathcal{I}_i^{\text{central}}(t)
=
(\mathcal{M}_i,\mathcal{S}_i,\mathcal{O}_j)
$
where $\mathcal{M}_i=\mathcal{M}$, and $t_s=t_0$ and $t_r=t_{max}$, where $\{t_0,...,t_{max}\}$ denotes the entire time-interval required for the agent to traverse from its start to goal positions. Note that the central coordinator under this setting requires the information $O_j, j \ne i$ of all the other agents, including the neighboring agents.

In contrast, \emph{distributed} algorithms require each 
agent to make decisions based solely on {locally available} 
information, leading to
$
\mathcal{I}_i^{\text{distributed}}(t)
=
\bigl(
\mathcal{M}_i^k, \mathcal{S}_i, \mathcal{O}_j
\bigr),
$
where $\mathcal{M}_i$ may be a $k \times k$ subgrid of the map, often referred to as the \emph{FOV}, $\mathcal{S}_i$ is the agent's self information and $\mathcal{O}_j$, with $j \in \mathcal{N}_i$ is the neighboring agents' information which is either sensed or exchanged.

On the other hand, in decentralized algorithms, each agent only has access to the environment information and self-information, using which they independently plan out paths. So the information in these strategies can be denoted as
$
\mathcal{I}_i^{\text{decentral}}(t)
=
\bigl(
\mathcal{M}_i^k, \mathcal{S}_i
\bigr).
$

Finally, \emph{hybrid} algorithms combine global and local 
reasoning: a central module resolves large-scale couplings or 
computes partial plans, while agents use {local information }
for real-time coordination, yielding
$
\mathcal{I}_i^{\text{hybrid}}(t)
=
\bigl(
\mathcal{M}_i,\, \mathcal{S}_i, \mathcal{O}_j, \mathcal{E}_{ij}
\bigr),
$
where additional information $\mathcal{O}_j$ related to other agents is shared to an agent only under the occurrence of specific \emph{events} $\mathcal{E}_{ij}$, like collisions or deadlocks.
Thus, although all MAPF settings require the same 
foundational information, they differ fundamentally in how 
much of it is shared, how widely it is distributed, and which 
components are used to compute safe and coordinated 
multi-agent behavior. To illustrate this, the information {exchange} across various categories of well known MAPF algorithms is presented in Table~\ref{tab:information_mapf}.
}

\subsection{Proposed Approach for IC--MAPF}

To address this problem of information-centric MAPF, we treat collision resolution as a mechanism for controlling when additional information about other agents needs to be shared. 
At a high level, a hybrid four-stage strategy can be built. In S1 (Decentralized Path Planning), each agent independently computes a plan from its start to its goal using information about the map and static obstacles. In S2 (Centralized Collision Detection), a coordinating module examines the collection of agent plans over time and identifies conflicts (vertex or edge conflicts). In S3 (Collision Control), a central control policy interprets each detected conflict and specifies a set of constraints to resolve the conflict. Finally, in S4 (Replanning), one of the agents involved in the conflict replans its path following the constraints from S3.

By varying what constraints are set for replanning, we obtain a family of tiered replanning strategies (S4.1–S4.4) that trade off information usage against solution completeness. (S4.1) yield-based local coordination, where a single agent performs a short detour to a nearby parking location, waits, and then rejoins its original plan using only immediate (typically 1 to 3 hop distance) local occupancy information; (S4.2) static replanning, where a small segment of one agent’s plan is recomputed while treating a fixed set of conflict cells as forbidden and keeping all other plans unchanged; (S4.3) dynamic replanning, where a small segment is replanned while treating short prefixes of other agents’ plans as time-varying obstacles over a finite horizon; and (S4.4) local joint planning, where a tightly coupled subset of agents is replanned jointly over a bounded window while all remaining agents are treated as fixed reservations. Together, these tiers provide a controlled way to gradually increase the amount of information used for coordination only when simpler, lower inter-agent information-based repairs fail. We provide completeness and soundness properties of these individual strategies and the aggregated strategy we employed in this work in Appendix~\ref{apx:tiered-properties}. The proposed methodology of the four-stage MAPF pipeline is detailed below.

\paragraph{S1: Decentralized Path Planning} 
Each agent $a_k$ generates an independent trajectory
\begin{align}
    \rho_k =\pi_{\theta}(s_k,g_k) = (v^k_0,\dots,v^k_{\tau_k}),  
\end{align}
where  $v^k_0 = s_k$, $v^k_{\tau_k}=g_k$, and \(\pi_{\theta}\) is a parameterized RL policy trained to minimize path length and collision risk over a planning horizon \(H\).  No information about other agents is exchanged during this stage. Let $\tau_i$ denote the set of makespan of each agent and $\tau = \{\tau_1,\dots,\tau_n\}$ be the set of all such makespans. 

\paragraph{S2–S3: Centralized Collision Detection and Control} 

A central module takes as input the independent trajectories of all the agents \(\rho=\{\rho_1,\dots,\rho_n\}\) along with the makespan set $\tau$, and identifies all vertex and edge conflicts, using


\begin{align}
C(\rho,\tau)=\{(t_j,\Delta_c,A_c)\mid {}
(v^k_{t_j} = v^l_{t_j} := \Delta_c)
\;\lor\;\nonumber\\
((v^k_{t_j},v^k_{t_{j+1}}) = (v^l_{t_j+1},v^l_{t_{j}}) := \Delta_c)\} \\
\nonumber\\ A_c=\{\{a_k,\,a_l\}:\rho_k|\{t_j\} = \rho_l|\{t_j\} = \Delta_c \lor \rho_k|\nonumber\\\{(t_j,t_{j+1})\}=\rho_l|\{(t_j,t_{j+1})\}=\Delta_c&, \forall k\ne l
\},
\end{align}
where $\rho_k|\{t_j\}$ and $\rho_k|\{(t_j,t_{j+1})\}$ denotes the position of the $k^{th}$ agent at step $t_j$ and steps ($t_j$, $t_{j+1}$), respectively. 

For each conflict \(c=(t_j,\Delta_c,A_c)\in C(\rho,\tau)\), which occurs at timestep $t_j$, the controller issues an alert $\mathcal{A}$ defined as
\begin{align}\mathcal{A}(c) = \mu_{c_k} = (a_{c_k},t_{j-r},\Delta_c), r \geq j,
\end{align}
where policy $\mu_{c_k}$ is used to select an agent $a_{c_k} \in A_c$. Example policies for this agent selection is described in Section \ref{main: method_policy}. Once an alert is issued to an agent, the selected agent is prompted to perform a constrained replanning of its trajectory for a rollout window $\{t^{c_k}_{j-r},\dots,t^{c_k}_{j},\dots,\tau_{c_k}\}$, where $r$ is the rewind window by avoiding the collision set $\Delta_c$.

\paragraph{S4: Replanning}

Upon receiving a collision alert for conflict time $t_j$ and rewind parameter $r$, the selected agent \(a_{c_k}\) conceptually decomposes its original trajectory \(\rho_{c_k} = (v^{c_k}_0,\dots,v^{c_k}_{\tau_{c_k}})\) into a fixed prefix, a replanning window, and an untouched suffix. We denote the fixed prefix up to just before the rewind step as
\begin{align}
    \rho_{c_k|t_{j-r-1}}=(v^{c_k}_0,\dots,v^{c_k}_{t_{j-r-1}}).
\end{align}

All S4 operators then construct a new path segment $\rho_{c_k}'$ that replaces the original states on the interval $\{t_{j-r}, ...,t_{j+r}\}$.

\emph{(S4.0) Yield-based local coordination.}
As a first step, the controller attempts a lightweight yield maneuver. If the conflict set $A_c$ contains an agent that is already parked at its start or has reached and is waiting at its goal at time $t_j$, we mark this agent as \emph{anchored} and select it as the replanning agent \(a_{c_k}\); otherwise, S3 selects the replanning agent \(a_{c_k} \in A_c\) using a heuristic policy (e.g., farthest from goal). In both cases, the remaining agents in $A_c$ are treated as protected and keep their current plans. Starting from its rewind state $v^{c_k}_{t_{j-r}}$, \(a_{c_k}\) then executes a short-horizon yield: it moves to a nearby parking cell, waits while the protected agent(s) traverse the conflict region, and then returns to a state from which it can resume progress toward $g_{c_k}$. The parking cell is chosen by a local search that ensures the resulting detour is free of vertex and edge conflicts with all other agents over the considered window. If no such parking cell exists, or if the yield trajectory does not eliminate the conflict when re-simulated by the controller, the plan is abandoned, and the system falls back to try the next strategy.

For static obstacle avoidance \emph{(S4.1)}, the constraint set, $\Delta_c$ in Eq.~\ref{eq:replan_policy_constrained}, is fixed by the alert and encodes the static cells that must be avoided. The ``bounded'' aspect comes from limiting how much of the trajectory is recomputed in time: we select a sub-initial point $v^{c_k}_{t_{j-r}}$ and a sub-goal (either the true goal $g_{c_k}$ or $v^{c_k}_{t_{j+r}}$), and only replan this suffix under the fixed constraint. The segment of the new path is generated as:
\begin{align}
    \rho_{c_k}' = \pi_{\theta}(v^{c_k}_{t_{j-r}}, v^{c_k}_{t_{j+r}} \mid v^{c_k}_{t_i} \notin \Delta_c, \forall t \in \{t_{j-r},...,t_{j+r}\}). \label{eq:replan_policy_constrained}
\end{align}

If bounded static replanning still fails to resolve the conflict, the controller may share richer information and trigger dynamic obstacle avoidance \emph{(S4.2)}, where the replanning must account for the predicted movements of other agents involved in the collision. In this case, the RL policy is conditioned on the sub-paths of the conflicting agents as illustrated by the constraints in Eq.~\ref{eq:replan_policy_dynamic}
\begin{align}
    \rho_{c_k}' = \pi_{\theta}(v^{c_k}_{t_{j-r}}, v^{c_k}_{t_{j+r}} \mid 
     v^{c_k}_{t} \neq v^{c_l}_{t}, v^{c_l}_{t} \in \rho_{c_l},\nonumber\\ \forall l \neq k, a_{c_l} \in A_c, \forall t \in \{t_{j-r},..., t_{j+r}\}).
    \label{eq:replan_policy_dynamic}
\end{align}

The information guiding the replanning--whether it constitutes minimal details such as static obstacle constraints (Eq.~\ref{eq:replan_policy_constrained}) or more detailed sub-path information about colliding agents' paths treated as dynamic obstacles (Eq.~\ref{eq:replan_policy_dynamic})--is integrated into the RL agent's decision-making process (by modifying its state representation, more details in Section~\ref{main: method_policy}) to guide it towards a conflict-free solution.

In all cases, the agent's new complete trajectory $\rho^{new}_{c_k}$ is formed by concatenating the initial segment with the newly planned one, given by

\begin{align}
    \rho^{new}_{c_k}
    = \rho_{c_k|t_{j-r-1}} \,\Vert\, \rho_{c_k}' \,\Vert\, \rho_{c_k|t_{j+r+1}}.
    \label{eq:new_concat_new_path}
\end{align}

Each updated trajectory is submitted to the central controller. The iterative cycle of detection (S2), control (S3), and decentralized replanning (S4) continues until no further conflicts can be resolved by single-agent strategies or a global termination condition is reached.

In a small fraction of cases, even dynamic single-agent replanning leaves a persistent local conflict among a tightly coupled subset of agents. To address these situations, the controller invokes a local joint planner \emph{(S4.3)} over a subset $A_c^{J} \subseteq A_c$ and a bounded time window around the conflict. Let $\{t_{j-r_J},..., t_{j+r_J}\}$ denote a short horizon centered at $t_j$. For each agent $a_k \in A_c^{J}$, we treat $v^{k}_{t_{j-r_J}}$ as its joint-planning start state and $v^{k}_{t_{j+r_J}}$ as a temporary sub-goal, and we solve a small-horizon problem in the joint configuration space of $A_c^{J}$ over this interval while treating all other agents as fixed reservations. The resulting joint segment $\{\rho_k^{\mathrm{joint}}\}_{a_k \in A_c^{J}}$ is spliced into the original trajectories between $t_{j-r_J}$ and $t_{j+r_J}$, analogous to Eq.~\ref{eq:new_concat_new_path}.

If none of the above strategies can successfully remove a conflict, the corresponding agent is temporarily \emph{deferred}: it is parked (typically at its start) and excluded from further replanning during the current phase. Once the non-deferred agents have reached a conflict-free configuration up to the current makespan, a second phase is initiated in which the deferred agents are reintroduced and planned using the same tiered strategies. This two-phase design specifically targets deadlock-like situations by decoupling stubborn local clusters and allowing them to be resolved after the rest of the system has stabilized.

\subsection{Execution Flow}
Our hybrid MAPF framework manages a precise flow of information, which is central to our goal of reducing inter–agent communication while preserving solution quality. The process begins with fully decentralized agent planning (S1), where each RL-driven agent uses only its local information to compute an intended path $\rho_k$ and submits this plan to a central coordinator. The coordinator, leveraging its global view of all submitted paths, performs collision detection (S2) to identify the set of potential conflicts $C(\rho,\tau)$. For each conflict, the control module (S3) issues a targeted alert $\mathcal{A}(c)$: it selects a replanning agent $a_{c_k}$, chooses a rewind window $\{t_{j-r},.., t_{j+r}\}$, and specifies which tiered strategy should be invoked.

In S4, the selected agent first attempts to resolve the conflict using minimal information through a local yield maneuver (S4.0), which relies only on occupancy checks and a nearby parking cell. If the yield is not applicable or fails, S3 gradually increases the information budget: it provides a static constraint set derived from the conflict (S4.1), then, if needed, short sub-paths of conflicting agents are treated as dynamic obstacles over the same window (S4.2). When these single-agent repairs are insufficient, a local joint A* over a small group of agents and a bounded horizon is invoked (S4.3), with all remaining agents treated as fixed reservations. Agents that still cannot be repaired are temporarily deferred and handled in a second phase with relaxed attempt limits to break deadlocks.

As shown in Appendix~\ref{apx:tiered-properties}, each tiered strategy in S4 is \textbf{sound}: whenever it returns a modified plan, the resulting joint execution remains collision-free over its replanned window. Moreover, under standard MAPF assumptions on finite grids with individually feasible start-goal pairs, the aggregated strategy with the deferral mechanism yields a \textbf{complete} repair procedure for our setting: every resolvable conflict is eventually eliminated, or else identified as belonging to an irreducible deadlock component among deferred agents. This tiered execution loop
ensures that additional information is only shared on demand, allowing our framework to maintain feasibility while operating with significantly reduced information exchange.

\subsection{Policy Representation \label{main: method_policy}}

\paragraph{(S1 \& S4) Decentralized Path Planning:}
Our framework employs an RL-based decentralized planner for dynamic replanning \emph{S4.2} that incorporates collision awareness directly into the agent’s observation, enabling effective path planning in dynamic multi-agent environments.

\textbf{Observation Space:}
Each agent observes a tensor $\mathbf{s} \in \mathbb{R}^{H \times W \times 4}$, where the channels encode: (a) a binary \emph{ObstacleMap} for static obstacles, (b) an \emph{AgentMap} for the agent’s current position, (c) a \emph{GoalMap} denoting the goal location, and (d) an \emph{AlertMask}, initially zero and updated online to reflect collision alerts generated by the centralized detection module. The observation is augmented with low-dimensional features consisting of a unit vector toward the goal and the Euclidean distance to the goal.

\textbf{Action Space:}
The agent operates in a discrete action space $\mathcal{A}=\{0,1,2,3,4\}$, corresponding to movements in the four cardinal directions and a \textsc{WAIT} action.

\textbf{Reward Structure:}
The reward function, following standard in learning-based approaches, encourages efficient, collision-free navigation. Agents receive a reward of +20 upon reaching the goal and are penalized for obstacle collisions (-3), timeouts (-2), each timestep (-0.02), and waiting actions (-0.1). An additional penalty of -0.05 is applied when the agent is near a dynamic obstacle.

\paragraph{(S2) Rule-Based Collision Detection:}
The collision detection module (S2) functions as a deterministic, rule-based system. It takes the set of all current agent trajectories \(\rho = \{\rho_1,\dots,\rho_n\}\) as input. For each timestep $t=\{0,1,...,T_{M}\}$, 
where $T_M$ is maximum makespan, the module systematically scans for vertex and edge conflicts, and reports these to the S3 control module for resolution. The detection process for all conflicts is computationally efficient, with a time complexity of \(O\bigl(\sum_k|\rho_k|\bigr)\) per cycle, linear in the sum of all agent path lengths.

\paragraph{(S3) Heuristic-Based Collision Avoidance Control:}
Upon notification of a conflict $c=(t,v,A_c)$ from S2, the S3 control module formulates and issues an alert $\mathcal{A}(c)$ to a selected agent. This involves choosing an agent $a_{c_k} \in A_c$ to replan, determining its replan interval $\{t_{j-r}, t_{j+r}\}$ by selecting an appropriate rewind value $r$, and specifying the replanning approach. We consider three agent selection policies (from $A_c$): (i) \emph{Random} choice (i.e., \(a_{c_k} \sim \text{UniformRandom}(A_c)\)), (ii) selecting the agent \emph{Farthest} from its goal ($g_a$) based on Manhattan distance $d_{\mathrm{Manh}}(v^a_{t_{j-r}},g_a)$, or (iii) identifying the agent with the \emph{Fewest Future Collisions (FFC)}, i.e., \(a_{c_k} = \arg\min_{a\in A_c}\;\bigl|\{\tilde c \mid \tilde c \in C(\rho,\tau),\,a\in A_{\tilde c}\}\bigr|\). In our work, we present the results with \emph{FFC} agent selection policy. 

\section{Experimental Setup \label{main: expt}}
In this section, we first describe how we train the per‐agent navigation policy used in stage S1, then benchmark dataset and planner details used to evaluate stages S2–S4, and define the performance metrics used to evaluate. We additionally provide details on the hardware experiments performed.

\paragraph{Training Procedure}
We train a parametrized Q-network $Q_{\theta}(s,a)$ using Double DQN \cite{van2016deep} with prioritized experience replay (PER) and $\varepsilon$-greedy exploration. The policy is trained to reach a specified goal while avoiding both static and dynamic obstacles. Dynamic obstacles follow valid precomputed trajectories with hidden goals, enabling the simulation of online collision alerts during inference from the S3 stage. Training is conducted for $30{,}000$ episodes on an $11\times11$ maze, with each episode capped at $T_{\max}=50$ steps. Environmental difficulty is increased through a curriculum: the first 500 episodes use static obstacle density $\rho_s=0.10$ with no dynamic obstacles; episodes 500--2999 use $\rho_s=0.10$ with one dynamic obstacle; episodes 3000--5999 use $\rho_s=0.20$ with two dynamic obstacles; and all remaining episodes use $\rho_s=0.30$ with four dynamic obstacles.

Observations are normalized to zero mean and unit variance and stored as transitions $(s_t,a_t,r_t,s_{t+1})$ in a PER buffer of size $10^6$. Mini-batches of size 128 are sampled for learning. Action selection follows an $\varepsilon$-greedy policy, with $\varepsilon$ decayed from $\varepsilon_0=1.0$ to $0.01$ according to $\varepsilon_{t+1}=\max(0.01,0.999\,\varepsilon_t)$. At all times, action selection and target computation are constrained by a validity mask $m_t\in\{0,1\}^{|\mathcal{A}|}$, ensuring that only feasible actions are considered \cite{Damani2021PRIMAL2}. Network parameters are optimized using Adam with learning rate $\alpha=3\times10^{-4}$ and discount factor $\gamma=0.97$. A separate target network with parameters $\theta^{-}$ is updated every 300 steps. The Double DQN target is computed as $y_t=r_t+\gamma\,Q_{\theta^{-}}(s_{t+1},\arg\max_{a'}Q_{\theta}(s_{t+1},a')\ \text{s.t.}\ m_t(a')=1)$, and the temporal-difference error as $\delta_t=y_t-Q_{\theta}(s_t,a_t)$. Training minimizes the PER-weighted Bellman loss $L(\theta)=\mathbb{E}_{i\sim p(i)}[w_i\,\delta_i^2]$, where $p(i)\propto|\delta_i|^\alpha$ and $w_i$ are importance-sampling weights.


\vspace{-0.2em}

\paragraph{Simulation: Evaluation Scenarios and Baselines}
We evaluate IC--MAPF on standard benchmark maps from the \emph{Moving AI Lab} dataset \cite{stern2019mapf}, including random grids ($32\times32$, $64\times64$ with 20\% obstacles) and structured maps (\texttt{den312d} and \texttt{warehouse}). Agent counts range from 10--128 on random maps and 8--128 on structured maps. For each configuration, we generate 20 random instances and report aggregated results in Table~\ref{tab:search-vs-learning}. We compare IC--MAPF against representative search- and learning-based baselines: CBS, DCC, SCRIMP, and EPH. IC--MAPF is given a 10-minute limit per instance (Python implementation), CBS is capped at 120 seconds, and learning-based solvers are run with step limits (128 for random maps, 256 for \texttt{den312d}/\texttt{warehouse}) and a 5-minute wall-clock bound. Section~\ref{main: results} presents comparative results highlighting IC--MAPF’s scalability, solution quality, and communication efficiency.

\vspace{-0.2em}
\paragraph{Performance Criteria}
We evaluate our MAPF framework using standard metrics. \emph{Success Rate (SR)} is the proportion of instances solved within defined time limits. For successful instances, \emph{Makespan (MS)} measures the time until the last agent reaches its goal. 
To quantify information sharing, we define an abstract \textbf{Information Unit (IU)} as the data required to represent one agent’s state (e.g., its coordinates) at a single timestep. For learning-based methods, IU is estimated by counting the number of other agents observed within an agent’s field of view at each timestep, aggregated over all agents and time.  This unified abstraction enables direct comparison of information load across algorithms. For IC--MAPF, each cell-level constraint issued in S3 for S4 replanning counts as one IU. For learning-based methods, IU reflects the other-agent information visible within each agent’s FOV, and we additionally count any inter-agent communication (as in DCC and EPH). We omit IU for CBS because it uses full global information.

\vspace{-0.2em}


\paragraph{Hardware Experiments: TurtleBot4}
We evaluate our approach on five TurtleBot4 robots operating in a $6{\times}6$ indoor grid. Five problem instances are executed using a ROS~2 discovery server to ensure stable communication. Robots interface with the central controller only through a shared synchronization channel, with all other ROS topics kept isolated and no onboard perception used. At the start, each robot resets its pose via the ROS~2 \texttt{reset\_pos} service. Execution proceeds in strict lockstep: at each step, a robot (i) performs the required in-place rotation and acknowledges completion, and then (ii) executes a fixed $0.45$\,m forward motion and acknowledges again. The controller advances the system only when all robots have completed both actions, ensuring fully synchronized multi-robot execution. Details on test maps, setup images, and robot execution metrics are provided in Appendix Figures~\ref{app-fig:robo_grid_setup}--\ref{app-fig:tb4_setup} and Tables~\ref{app-tab:tb4_comm_summary_comm}--\ref{app-tab:tb4_comm_summary_quality}.

\begin{table*}[t]
  \centering
  \scriptsize
  \setlength{\tabcolsep}{3pt}
  \renewcommand{\arraystretch}{1.1}

  \begin{tabular}{c c c | *{14}{r}}
    \toprule
    & & & \multicolumn{2}{c}{Search-based} & \multicolumn{3}{c}{Hybrid} & \multicolumn{9}{c}{Learning-based Solvers} \\
    \cmidrule(lr){4-5} \cmidrule(lr){6-8} \cmidrule(lr){9-17}
    Map & & $m$ &
    \multicolumn{2}{c}{CBS} &
    \multicolumn{3}{c}{IC--MAPF} &
    \multicolumn{3}{c}{SCRIMP} &
    \multicolumn{3}{c}{DCC} &
    \multicolumn{3}{c}{EPH} \\
    & & &
    SR$\uparrow$ & MS$\downarrow$ &
    SR$\uparrow$ & MS$\downarrow$ & IU$\downarrow$ &
    SR$\uparrow$ & MS$\downarrow$ & IU$\downarrow$ &
    SR$\uparrow$ & MS$\downarrow$ & IU$\downarrow$ &
    SR$\uparrow$ & MS$\downarrow$ & IU$\downarrow$ \\
    \midrule

    \multirow{5}{*}{\includegraphics[height=1.5cm]{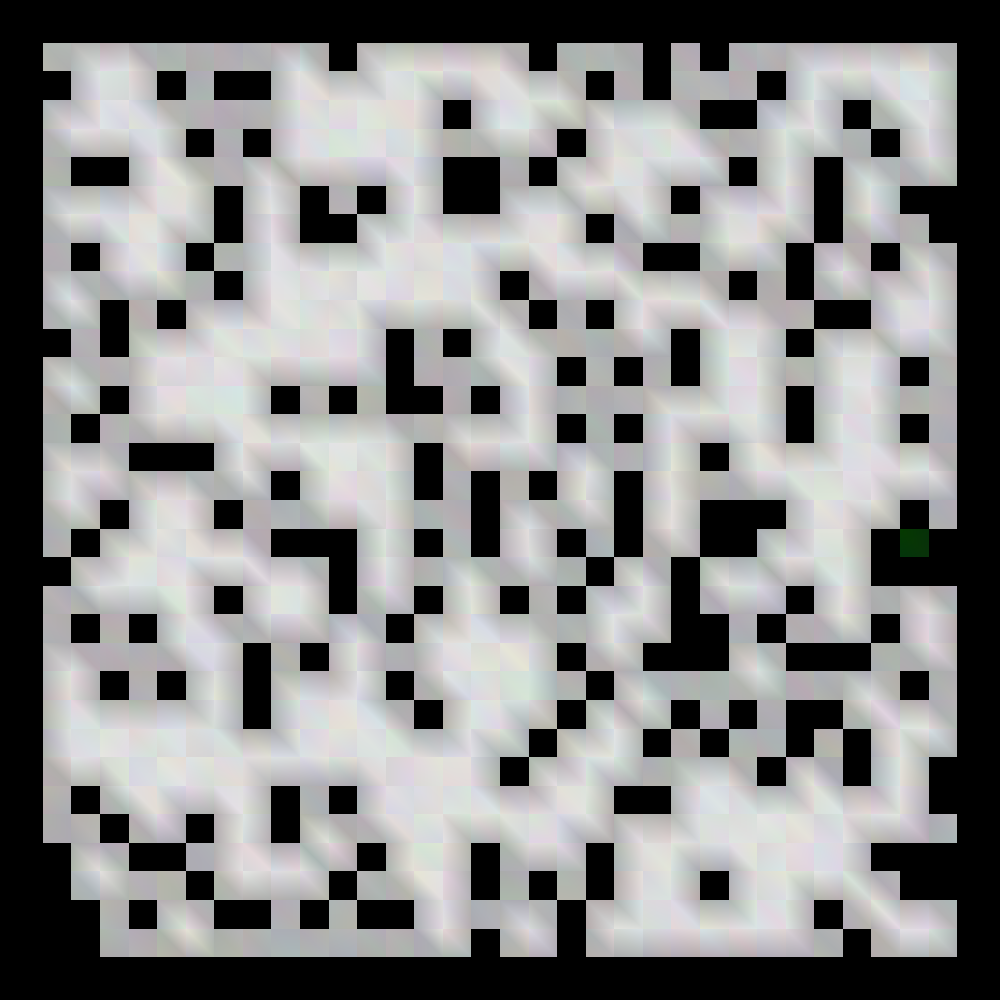}} &
    \multirow{5}{*}{\rotatebox{90}{\shortstack{\texttt{random}\\\texttt{32$\times$32-20}}}} &
    10-30 & \textbf{100} & \textbf{47.6} & \textbf{100} & 54.0 & \textbf{29.6} & \textbf{100} & 50.3 & 19x & \textbf{100} & 55.5 & 11x & 98 & 57.4 & 9.0x \\
     & &
    40-60 & 65 & 85.3 & \textbf{100} & 98.9 & \textbf{161.5} & 95 & \textbf{62.1} & 12x & 85 & 82.1 & 23x & 97 & 77.4 & 22x \\
     & &
    70-90 & 0 & - & \textbf{98} & \textbf{73.4} & \textbf{1824.2} & 77 & 83.8 & 2.6x & 50 & 111.3 & 10x & 53 & 114.5 & 12x \\
     & &
    100 & 0 & - & \textbf{85} & \textbf{76.5} & \textbf{3553.8} & 60 & 97.2 & 2.0x & 40 & 117.5 & 11x & 50 & 121.0 & 12x \\
     & &
    128 & 0 & - & 0 & - & - & \textbf{70} & \textbf{96.6} & - & 0 & 128.0 & - & 5 & 127.5 & - \\
    \midrule

    \multirow{5}{*}{\includegraphics[height=1.5cm]{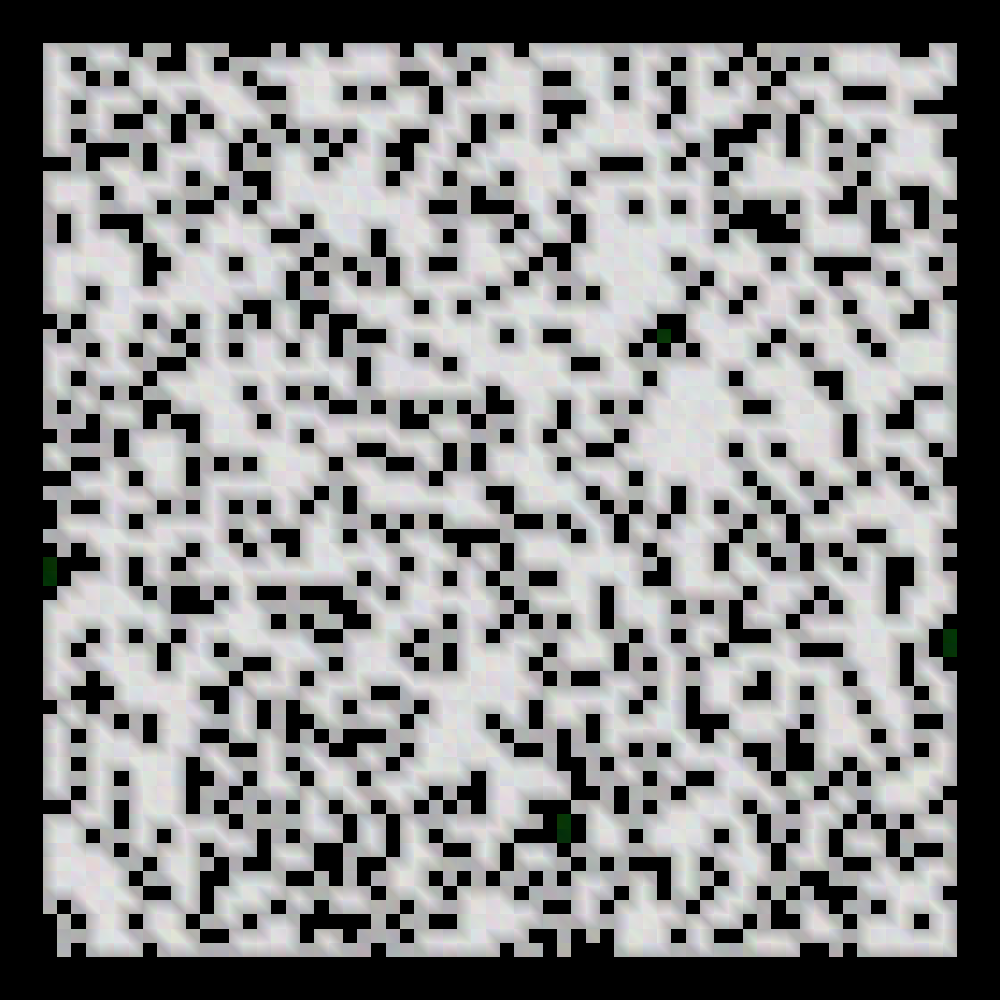}} &
    \multirow{5}{*}{\rotatebox{90}{\shortstack{\texttt{random}\\\texttt{64$\times$64-20}}}} &
    10-30 & 85 & 106.3 & \textbf{100} & 113.9 & \textbf{10.2} & 23 & 121.9 & 137x & 97 & 105.9 & 7.9x & 98 & \textbf{104.1} & 6.5x \\
     & &
    40-60 & 42 & 130.0 & \textbf{100} & 125.6 & \textbf{88.5} & 0 & 128.0 & - & 83 & 116.8 & 9.5x & 98 & \textbf{114.0} & 7.3x \\
     & &
    70-90 & 0 & - & \textbf{98} & 129.1 & \textbf{213.5} & 0 & 128.0 & - & 43 & 125.0 & 14x & 85 & \textbf{121.1} & 11x \\
     & &
    100 & 0 & - & \textbf{90} & 130.1 & \textbf{462.1} & 0 & 128.0 & - & 25 & 127.2 & 13x & 45 & \textbf{125.3} & 11x \\
     & &
    128 & 0 & - & 5 & 132.0 & \textbf{607.6} & 0 & 128.0 & - & 0 & 128.0 & - & \textbf{15} & \textbf{127.7} & 20x \\
    \midrule

    \multirow{5}{*}{\includegraphics[height=1.5cm]{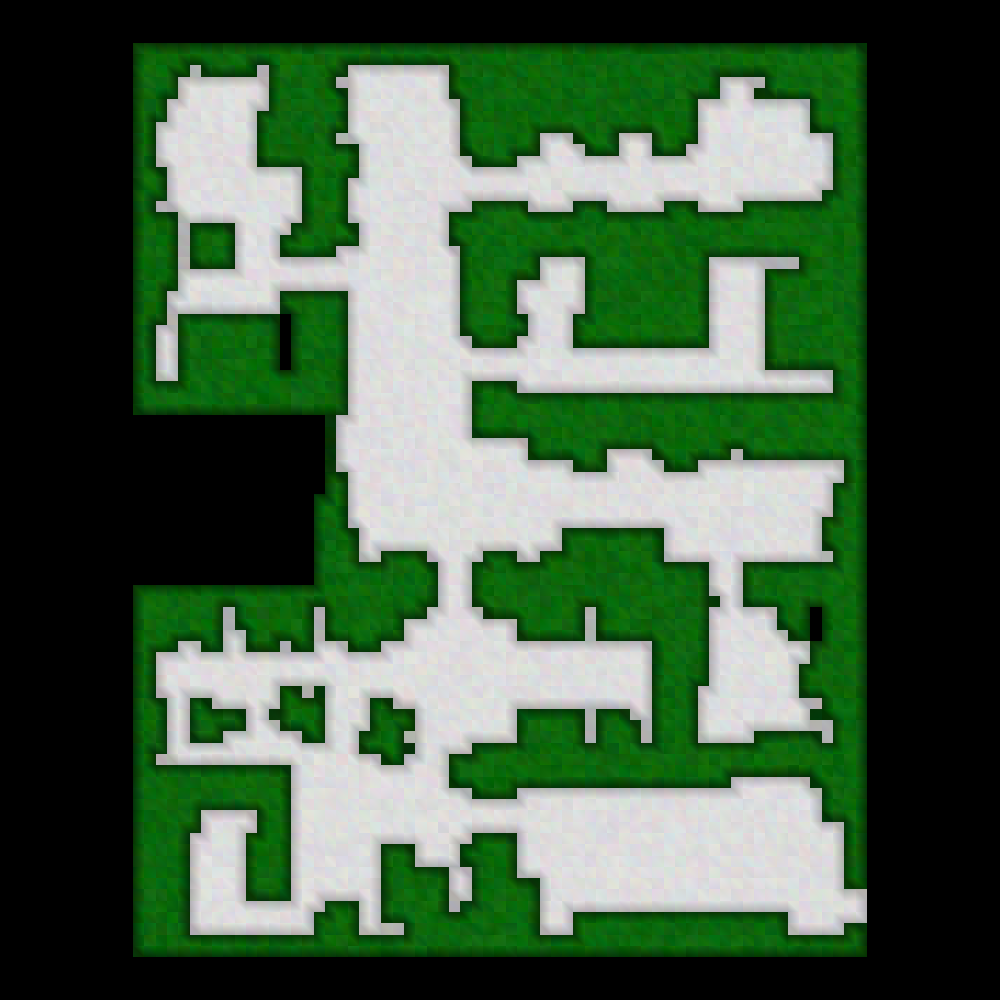}} &
    \multirow{5}{*}{\rotatebox{90}{\shortstack{\texttt{den312d}\\\texttt{65$\times$81}}}} &
    8 & \textbf{100} & \textbf{97.2} & \textbf{100} & 115.3 & \textbf{3.6} & 20 & 226.2 & 317x & \textbf{100} & 114.2 & 6.2x & \textbf{100} & 111.9 & 3.6x \\
     & &
    16 & 85 & 128.7 & \textbf{100} & 124.0 & \textbf{21.8} & 0 & 256.0 & - & \textbf{100} & 130.7 & 8.8x & \textbf{100} & \textbf{122.2} & 5.9x \\
     & &
    32 & 20 & 148.0 & \textbf{100} & 136.7 & \textbf{105.2} & 0 & 256.0 & - & 85 & 163.5 & 11x & \textbf{100} & \textbf{127.0} & 6.0x \\
     & &
    64 & 0 & - & \textbf{100} & 153.1 & \textbf{759.9} & 0 & 256.0 & - & 90 & 190.7 & 11x & \textbf{100} & \textbf{146.1} & 8.8x \\
     & &
    128 & 0 & - & 0 & - & - & 0 & 256.0 & - & 5 & 254.8 & - & \textbf{90} & \textbf{207.7} & - \\
    \midrule

    \multirow{5}{*}{\includegraphics[height=1.5cm]{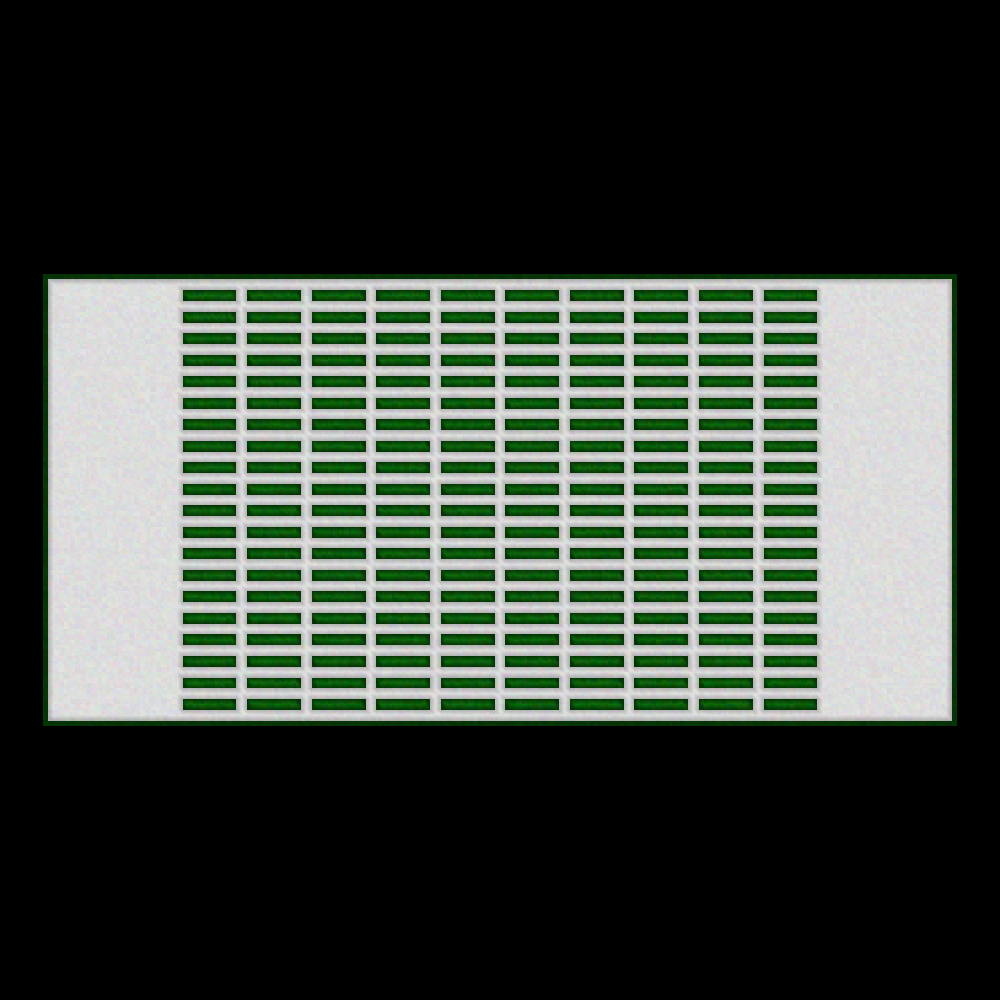}} &
    \multirow{5}{*}{\rotatebox{90}{\shortstack{\texttt{warehouse}\\\texttt{161$\times$63}}}} &
    8 & \textbf{100} & \textbf{175.7} & \textbf{100} & 192.8 & \textbf{0.8} & 0 & 256.0 & - & \textbf{100} & 191.8 & 5.8x & \textbf{100} & 191.6 & 2.3x \\
     & &
    16 & \textbf{100} & \textbf{196.2} & \textbf{100} & 215.1 & \textbf{1.0} & 0 & 256.0 & - & \textbf{100} & 213.3 & 16x & \textbf{100} & 212.9 & 11x \\
     & &
    32 & 15 & 221.3 & \textbf{100} & 220.3 & \textbf{10.7} & 0 & 256.0 & - & \textbf{100} & 219.2 & 8.3x & \textbf{100} & \textbf{218.1} & 5.4x \\
     & &
    64 & 0 & - & \textbf{100} & \textbf{227.6} & \textbf{23.6} & 0 & 256.0 & - & \textbf{100} & 227.6 & 15x & \textbf{100} & 227.8 & 11x \\
     & &
    128 & 0 & - & \textbf{100} & \textbf{234.7} & \textbf{139.1} & 0 & 256.0 & - & 95 & 236.8 & 14x & 90 & 239.1 & 11x \\
    \bottomrule
  \end{tabular}

    \caption{Performance of search-based (CBS, IC--MAPF) and learning-based (DCC, SCRIMP, EPH) MAPF solvers. $m$, MS, SR, IU represents the number of agents, success rate, makespan and information units respectively. Values are averaged within agent count groups. IU shows information units as Nx multiplier relative to our IC--MAPF baseline. Lower MS$\downarrow$ and IU$\downarrow$, higher SR$\uparrow$ are better. The shaded columns correspond to IC--MAPF results, our proposed hybrid approach.}

  \label{tab:search-vs-learning}
\end{table*}

\section{Results and Discussion \label{main: results}}
Here we present an empirical study, evaluating our central hypothesis: \emph{a MAPF framework with strategically reduced information sharing can achieve robust performance.} We address our \textbf{RQs} concerning minimal information needs, and Table~\ref{tab:search-vs-learning} summarizes the comparative performance of IC--MAPF against representative search-based (CBS) and learning-based (DCC, SCRIMP, EPH) MAPF solvers across random grids and structured maps. Additional experimental results are presented in Appendix Section~\ref{app:add_results} and detailed results with runtime in Figure~\ref{fig:results_plot}.

\textbf{(RQ1):} \emph{Given the observation constraints of a decentralized setup, can an effective MAPF algorithm be created with one agent knowing nothing about other agents? If not, what information must it need at a minimum?}

\textbf{Ans.} \textbf{No}--A purely decentralized setting in which agents have no information about others is insufficient once interactions occur, as conflicts cannot be consistently resolved. However, our results show that effective MAPF does not require continuous observation or global plan sharing. Instead, solutions can be achieved with \emph{minimal, event-triggered information sharing} consistent with our hybrid information model $\mathcal{I}_i^{\text{hybrid}}(t)$, where additional other-agent information is revealed only upon conflict events.

Across unstructured random grids, IC--MAPF achieves high success rates while using orders of magnitude less information than learning-based baselines. On the \texttt{random-32$\times$32-20} map, IC--MAPF maintains $100\%$ SR up to 80 agents, with cumulative IU values ranging from 3.6 to 214.2. Under the same settings, SCRIMP requires 262--2823 IU and EPH requires 19.7--6155 IU, corresponding to approximately $3\times$ to over $30\times$ higher information usage, while DCC requires 26.1--6337.2 IU (about $7\times$--$30\times$ IC--MAPF over the same range). On the larger \texttt{random-64$\times$64-20} map, IC--MAPF sustains $100\%$ SR up to 80 agents with IU below 212, whereas SCRIMP fails beyond 30 agents and exceeds 6000 IU, and EPH exhibits reduced SR while using $6\times$ to $13\times$ more information than IC--MAPF; DCC similarly achieves moderate SR but uses 10.6--2947.4 IU, i.e., roughly $7\times$--$14\times$ more information. Similarly, on structured benchmarks, \texttt{den312d}, IC--MAPF achieves $100\%$ SR for 8--64 agents with IU between 3.6 and 759.9, while EPH requires up to 6722.3 IU to obtain comparable SR, and SCRIMP fails for 16 agents and above; DCC lies in between, using 22.6--8180.1 IU (approximately $6\times$--$11\times$ IC--MAPF for 8--64 agents). On the \texttt{warehouse} map, IC--MAPF solves all tested instances (8--128 agents) with IU between 0.8 and 139.1, whereas EPH requires 1.7--1502.4 IU to achieve similar SR, and DCC uses 4.3--2016.5 IU (about $5\times$--$16\times$ IC--MAPF), while SCRIMP achieves $0\%$ SR despite using several thousand IU. These results identify \emph{targeted, conflict-triggered sharing of short agent-path fragments over bounded horizons} as the minimal information required for effective MAPF, enabling reduced total information load of up to one to two orders of magnitude. 

\textbf{(RQ2):} \emph{How does the proposed hybrid method compare to leading alternative search- and learning-based approaches in terms of performance, solution quality, and scalability?}

\textbf{\emph{Ans.}} Table~\ref{tab:search-vs-learning} comparison highlights three key dimensions: success rate (SR), makespan (MS), and total information load (IU). Search-based solvers such as CBS rely on full joint information \(\mathcal{I}_i^{\text{central}}(t)\) and attempt to resolve all couplings through centralized reasoning. While optimal when successful, CBS fails to scale under the imposed time limits and does not solve most medium-to-large instances reported in the table. This illustrates the practical limitations of centralized planners as problem size grows. Learning-based distributed methods, including DCC, SCRIMP, and EPH, operate under \(\mathcal{I}_i^{\text{distributed}}(t)\), where each agent continuously accesses local observations of other agents within a field of view or via explicit message passing. These methods show strong performance in some dense scenarios (e.g., EPH on \texttt{warehouse}), but incur substantially higher information loads. In Table~\ref{tab:search-vs-learning}, their IU values are reported as multipliers relative to IC--MAPF, often ranging from one to two orders of magnitude higher due to continuous perception or communication.

In contrast, IC--MAPF adopts a hybrid execution strategy. 
This allows IC--MAPF to scale favorably across both random and structured maps, achieving high SR with competitive makespans while keeping information exchange sparse and bounded. The gradual escalation of information, from static constraints to short dynamic sub-paths, and finally to bounded joint replanning only when necessary, ensures that increased coordination cost is incurred only in genuinely hard cases. Overall, IC--MAPF occupies a middle ground between centralized optimal planners and fully distributed learning-based planners. 
These results demonstrate that carefully controlled, event-driven information sharing can provide a favorable trade-off between performance and communication complexity in MAPF.

\section{Conclusions \label{main: conclusion}}
In this paper, we introduced the problem of \emph{Information-centric MAPF} and proposed a hybrid framework, IC--MAPF, that significantly reduces the inter-agent information required for successful multi-agent coordination. We further defined a metric, \emph{Information Units} (IU), to quantify the amount of other-agent information consumed during planning. Across a range of challenging MAPF benchmarks, IC--MAPF achieves high success rates while reducing information usage by $2\times$ to $23\times$ compared to state-of-the-art communication-efficient baselines.

The broader implications of this work extend to privacy-aware autonomous systems, where reducing inter-agent information exchange is essential. By showing that multi-agent coordination is achievable with sparse, event-triggered information, our results support the development of scalable, bandwidth-efficient, and privacy-preserving multi-robot systems. Despite these promising results, our work has limitations: IC--MAPF may face challenges in extremely dense settings where conflicts are frequent, and its tiered strategy is currently hand-designed. As future work, we aim to explore learning-based mechanisms that automatically adapt the tier-selection policy and investigate reinforcement learning under explicit information constraints, enabling agents to optimize performance while minimizing IU usage. 

\bibliography{references}

@article{ma2021learning,
  title={Learning selective communication for multi-agent path finding},
  author={Ma, Ziyuan and Luo, Yudong and Pan, Jia},
  journal={IEEE Robotics and Automation Letters},
  volume={7},
  number={2},
  pages={1455--1462},
  year={2021},
  publisher={IEEE}
}

@inproceedings{mapf-empiricalhardness-bluesky,
  title={Empirical Hardness in Multi-Agent Pathfinding: Research Challenges and Opportunities},
  author={Jingyao Ren and Ewing Eric and T. K. Satish Kumar and Sven Koenig and Nora Ayanian},
  booktitle={Blue Sky paper at 24th International Conference on Autonomous Agents and Multiagent Systems},
  year={2025}
}

@inproceedings{boyarski2015icbs,
  title={Icbs: The improved conflict-based search algorithm for multi-agent pathfinding},
  author={Boyarski, Eli and Felner, Ariel and Stern, Roni and Sharon, Guni and Betzalel, Oded and Tolpin, David and Shimony, Eyal},
  booktitle={Proceedings of the International Symposium on Combinatorial Search},
  volume={6},
  number={1},
  pages={223--225},
  year={2015}
}

@article{sartoretti2019primal,
  title={PRIMAL: Pathfinding via Reinforcement and Imitation Multi-Agent Learning},
  author={Sartoretti, G. and others},
  journal={IEEE Robotics and Automation Letters},
  volume={4},
  number={3},
  pages={2559--2566},
  year={2019}
}

@inproceedings{He2016CVPR,
  title     = {Deep Residual Learning for Image Recognition},
  author    = {He, Kaiming and Zhang, Xiangyu and Ren, Shaoqing and Sun, Jian},
  booktitle = {Proceedings of the IEEE Conference on Computer Vision and Pattern Recognition},
  pages     = {770--778},
  year      = {2016},
}

@inproceedings{van2016deep,
  title={Deep reinforcement learning with double q-learning},
  author={Van Hasselt, Hado and Guez, Arthur and Silver, David},
  booktitle={Proceedings of the AAAI conference on artificial intelligence},
  volume={30},
  number={1},
  year={2016}
}

@inproceedings{ma2021distributed,
  title={Distributed heuristic multi-agent path finding with communication},
  author={Ma, Ziyuan and Luo, Yudong and Ma, Hang},
  booktitle={2021 IEEE International Conference on Robotics and Automation (ICRA)},
  pages={8699--8705},
  year={2021},
  organization={IEEE}
}

@inproceedings{stern2019multi,
  title={Multi-agent pathfinding: Definitions, variants, and benchmarks},
  author={Stern, Roni and Sturtevant, Nathan and Felner, Ariel and Koenig, Sven and Ma, Hang and Walker, Thayne and Li, Jiaoyang and Atzmon, Dor and Cohen, Liron and Kumar, TK and others},
  booktitle={Proceedings of the International Symposium on Combinatorial Search},
  volume={10},
  number={1},
  pages={151--158},
  year={2019}
}

@article{stern2019mapf,
  title={Multi-Agent Pathfinding: Definitions, Variants, and Benchmarks},
  author={Roni Stern and Nathan R. Sturtevant and Ariel Felner and Sven Koenig and Hang Ma and Thayne T. Walker and Jiaoyang Li and Dor Atzmon and Liron Cohen and T. K. Satish Kumar and Eli Boyarski and Roman Bartak},
  journal={Symposium on Combinatorial Search (SoCS)},
  year={2019},
  pages={151--158}
}

@inproceedings{wang2023scrimp,
  title={Scrimp: Scalable communication for reinforcement-and imitation-learning-based multi-agent pathfinding},
  author={Wang, Yutong and Xiang, Bairan and Huang, Shinan and Sartoretti, Guillaume},
  booktitle={2023 IEEE/RSJ International Conference on Intelligent Robots and Systems (IROS)},
  pages={9301--9308},
  year={2023},
  organization={IEEE}
}

@inproceedings{WagnerChoset2011,
  title        = {M*: A Complete Multirobot Path Planning Algorithm with Performance Bounds},
  author       = {Wagner, Glenn and Choset, Howie},
  booktitle    = {Proceedings of the IEEE/RSJ International Conference on Intelligent Robots and Systems (IROS)},
  pages        = {3260--3267},
  year         = {2011},
}

@article{Sharon2013,
  title        = {The Increasing Cost Tree Search for Optimal Multi‑Agent Pathfinding},
  author       = {Sharon, Guni and Stern, Roni and Goldenberg, Meir and Felner, Ariel},
  journal      = {Artificial Intelligence},
  volume       = {195},
  number       = {C},
  pages        = {470--495},
  year         = {2013},
  doi          = {10.1016/j.artint.2012.11.006},
}

@article{Sharon2015,
  title        = {Conflict‑Based Search for Optimal Multi‑Agent Pathfinding},
  author       = {Sharon, Guni and Stern, Roni and Felner, Ariel and Sturtevant, Nathan R.},
  journal      = {Artificial Intelligence},
  volume       = {219},
  pages        = {40--66},
  year         = {2015},
  doi          = {10.1016/j.artint.2014.11.006},
}

@inproceedings{LiAAAI22,
  author       = {Li, Jiaoyang and Chen, Zhe and Harabor, Daniel and Stuckey, Peter J. and Koenig, Sven},
  title        = {MAPF‑LNS2: Fast Repairing for Multi‑Agent Path Finding via Large Neighborhood Search},
  booktitle    = {Proceedings of the AAAI Conference on Artificial Intelligence},
  volume       = {36},
  pages        = {10256--10265},
  year         = {2022},
  doi          = {10.1609/aaai.v36i9.21266},
}

@article{Damani2021PRIMAL2,
  title        = {PRIMAL2: Pathfinding via Reinforcement and Imitation Multi‑Agent Learning—Lifelong},
  author       = {Damani, Mehul and Luo, Zhiyao and Wenzel, Emerson and Sartoretti, Guillaume},
  journal      = {IEEE Robotics and Automation Letters},
  volume       = {6},
  number       = {2},
  pages        = {2666--2673},
  year         = {2021},
  doi          = {10.1109/LRA.2021.3062803},
}

@inproceedings{Skrynnik2024,
  title        = {Learn to Follow: Decentralized Lifelong Multi‑Agent Pathfinding via Planning and Learning},
  author       = {Skrynnik, Alexey and Andreychuk, Anton and Nesterova, Maria and Yakovlev, Konstantin and Panov, Aleksandr},
  booktitle    = {Proceedings of the AAAI Conference on Artificial Intelligence},
  volume       = {38},
  pages        = {17541--17549},
  year         = {2024},
}

@inproceedings{WangAAAI25,
  author       = {Wang, Yutong and Duhan, Tanishq and Li, Jiaoyang and Sartoretti, Guillaume Adrien},
  title        = {LNS2+RL: Combining Multi‑agent Reinforcement Learning with Large Neighborhood Search in Multi‑agent Path Finding},
  booktitle    = {Proceedings of the AAAI Conference on Artificial Intelligence},
  year         = {2025},
}

@article{he2025social,
  title={Social behavior as a key to learning-based multi-agent pathfinding dilemmas},
  author={He, Chengyang and Duhan, Tanishq and Tulsyan, Parth and Kim, Patrick and Sartoretti, Guillaume},
  journal={Artificial Intelligence},
  pages={104397},
  year={2025},
  publisher={Elsevier}
}

@article{okumura2022priority,
  title={Priority inheritance with backtracking for iterative multi-agent path finding},
  author={Okumura, Keisuke and Machida, Manao and D{\'e}fago, Xavier and Tamura, Yasumasa},
  journal={Artificial Intelligence},
  volume={310},
  pages={103752},
  year={2022},
  publisher={Elsevier}
}

@inproceedings{he2024alpha,
  title={Alpha: Attention-based long-horizon pathfinding in highly-structured areas},
  author={He, Chengyang and Yang, Tianze and Duhan, Tanishq and Wang, Yutong and Sartoretti, Guillaume},
  booktitle={2024 IEEE International Conference on Robotics and Automation (ICRA)},
  pages={14576--14582},
  year={2024},
  organization={IEEE}
}

@inproceedings{tang2024ensembling,
  title={Ensembling prioritized hybrid policies for multi-agent pathfinding},
  author={Tang, Huijie and Berto, Federico and Park, Jinkyoo},
  booktitle={2024 IEEE/RSJ International Conference on Intelligent Robots and Systems (IROS)},
  pages={8047--8054},
  year={2024},
  organization={IEEE}
}

@inproceedings{okumura2023lacam,
  title={Lacam: Search-based algorithm for quick multi-agent pathfinding},
  author={Okumura, Keisuke},
  booktitle={Proceedings of the AAAI Conference on Artificial Intelligence},
  volume={37},
  number={10},
  pages={11655--11662},
  year={2023}
}

@article{liao2025sigma,
  title={SIGMA: Sheaf-Informed Geometric Multi-Agent Pathfinding},
  author={Liao, Shuhao and Xia, Weihang and Cao, Yuhong and Dai, Weiheng and He, Chengyang and Wu, Wenjun and Sartoretti, Guillaume},
  journal={arXiv preprint arXiv:2502.06440},
  year={2025}
}

@inproceedings{SchulmanLevineAbbeelJordanMoritz2015,
  title     = {Trust Region Policy Optimization},
  author    = {Schulman, John and Levine, Sergey and Abbeel, Pieter and Jordan, Michael I. and Moritz, Philipp},
  booktitle = {Proceedings of the International Conference on Machine Learning},
  pages     = {1889--1897},
  publisher = {PMLR},
  month     = {June},
  year      = {2015},
  url       = {https://proceedings.mlr.press/v37/schulman15.html}
}

@article{SchulmanMoritzLevineJordanAbbeel2015,
  title   = {High-Dimensional Continuous Control Using Generalized Advantage Estimation},
  author  = {Schulman, John and Moritz, Philipp and Levine, Sergey and Jordan, Michael I. and Abbeel, Pieter},
  journal = {arXiv preprint arXiv:1506.02438},
  year    = {2015},
  url     = {https://arxiv.org/abs/1506.02438}
}

@article{SchulmanWolskiDhariwalRadfordKlimov2017,
  title   = {Proximal Policy Optimization Algorithms},
  author  = {Schulman, John and Wolski, Filip and Dhariwal, Prafulla and Radford, Alec and Klimov, Oleg},
  journal = {arXiv preprint arXiv:1707.06347},
  year    = {2017},
  url     = {https://arxiv.org/abs/1707.06347}
}

\newpage
\appendix
\section{Appendix}
\section*{Appendix Contents}
\addcontentsline{toc}{section}{Appendix Contents}

\noindent\ref{app:add_results}. Additional Results \dotfill \pageref{app:add_results}
\newline\hspace*{2em}Detailed Performance Table -- Table~\ref{tab:search-vs-learning} \dotfill \pageref{tab:search-vs-learning}
\newline\hspace*{2em}Strategy Usage Summary -- Figure~\ref{fig:strat_usage} \dotfill \pageref{fig:strat_usage}
\newline\hspace*{2em}Detailed Performance Plots -- Figure~\ref{fig:results_plot} \dotfill \pageref{fig:results_plot}


\vspace{0.5em}
\noindent\ref{apx:tiered-properties}. Properties of Tiered Replanning Strategies \dotfill \pageref{apx:tiered-properties}  

\begin{figure*}
    \centering
    \includegraphics[width=1\linewidth]{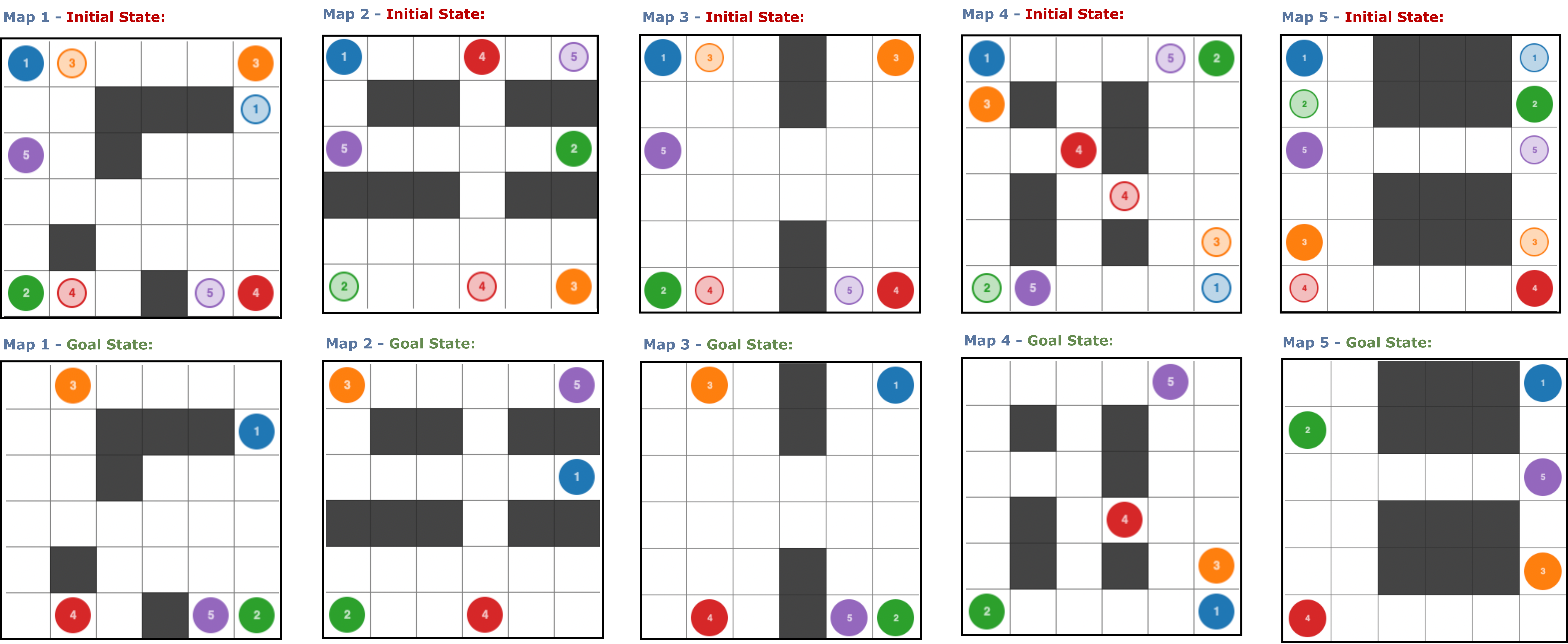}
    \caption{Five map configuration, initial state and goal state, used for testing our IC--MAPF algorithm on TurtleBot4s’}
    \label{app-fig:robo_grid_setup}
\end{figure*}

\begin{figure}
    \centering
    \includegraphics[width=1\linewidth]{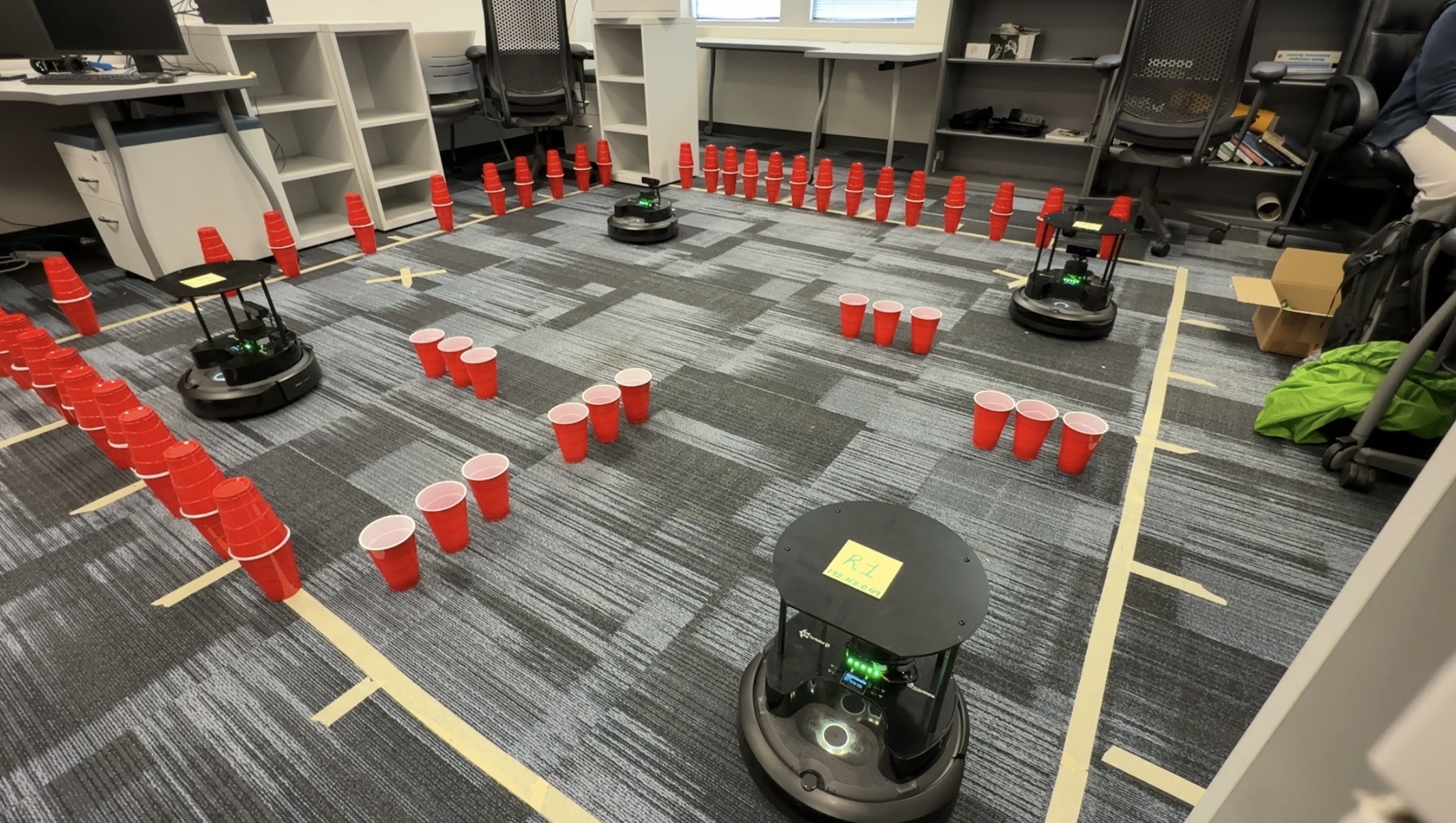}
    \caption{Indoor 6x6 grid setup with TurtleBot4'}
    \label{app-fig:tb4_setup}
\end{figure}

\vspace{0.5em}
\noindent\ref{app:robot_exp}. Hardware Results: TurtleBot4 \dotfill \pageref{app:robot_exp}
\newline\hspace*{2em}Robot Grid Configurations -- Figure~\ref{app-fig:robo_grid_setup} \dotfill \pageref{app-fig:robo_grid_setup}
\newline\hspace*{2em}Physical Setup Image -- Figure~\ref{app-fig:tb4_setup} \dotfill \pageref{app-fig:tb4_setup}
\newline\hspace*{2em}Communication Statistics -- Table~\ref{app-tab:tb4_comm_summary_comm} \dotfill \pageref{app-tab:tb4_comm_summary_comm}
\newline\hspace*{2em}Execution Quality Statistics -- Table~\ref{app-tab:tb4_comm_summary_quality} \dotfill \pageref{app-tab:tb4_comm_summary_quality}

\vspace{0.5em}
\noindent\ref{ap: lit}. Literature review \dotfill \pageref{ap: lit}
\newline\hspace*{4em}Literature Categorization Table \ref{tab:lit_comparison} \dotfill \pageref{tab:lit_comparison}

\begin{figure}[h!]
    \centering
    \includegraphics[width=1\linewidth]{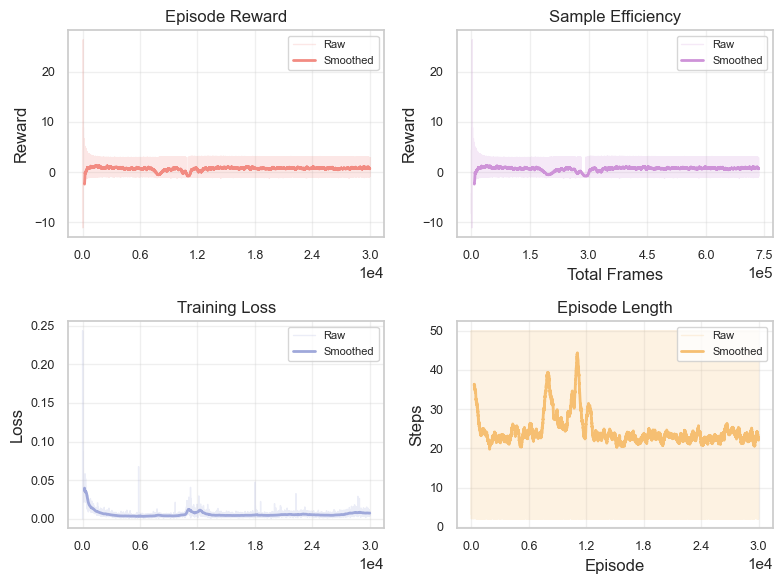}
    \caption{The plot illustrates the training performance of the \emph{Double Deep Q-Network (DDQN)} algorithm. Episode rewards (red), sample efficiency measured by rewards per total frames (purple), training loss (blue), and episode length (orange) are presented across episodes. Smoothed curves represent moving averages, enhancing the visibility of underlying performance trends.}
    \label{fig:ddqn-training-performance}
\end{figure}

\begin{figure}[h!]
    \centering
    \includegraphics[width=1\linewidth]{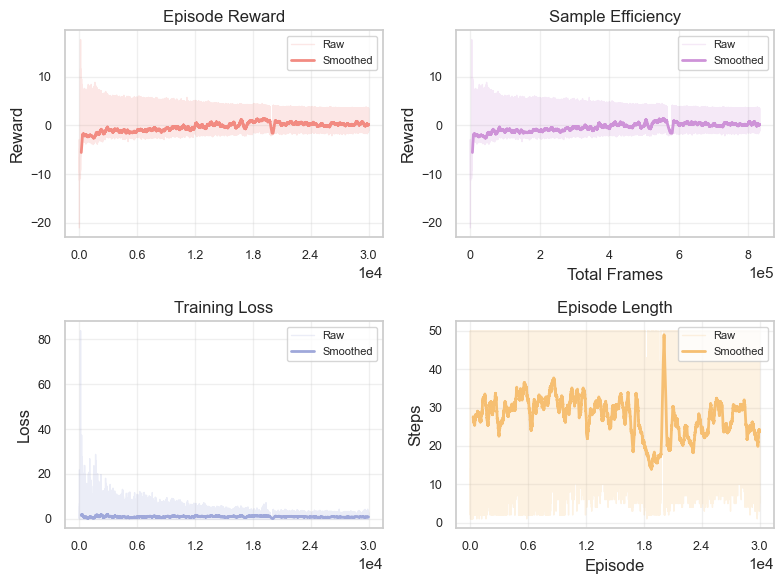}
    \caption{The plot illustrates the training performance of the \emph{Proximal Policy Optimization (PPO)} algorithm, capturing episode rewards (red). Episode rewards (red), sample efficiency measured by rewards per total frames (purple), training loss (blue), and episode length (orange) are presented across episodes. Smoothed curves represent moving averages, enhancing the visibility of underlying performance trends.}
    \label{fig:ppo-training-performance}
\end{figure}

\vspace{0.5em}
\noindent\ref{ap: train}. Training Methods \dotfill \pageref{ap: train}
\newline\hspace*{2em}\ref{ap: ppo}. PPO Training Procedure \dotfill \pageref{ap: ppo}
\newline\hspace*{2em}\ref{ap: network}. Neural Network Architecture \dotfill \pageref{ap: network}
\newline\hspace*{2em}\ref{ap: train_res}. Model Comparison \dotfill \pageref{ap: train_res}
\newline\hspace*{4em}DDQN training performance - Figure \ref{fig:ddqn-training-performance} \dotfill
\newline\hspace*{4em}PPO training performance - Figure \ref{fig:ppo-training-performance} \dotfill



\subsection{Additional Results} \label{app:add_results}

\begin{figure*}[!h]
    \centering
    \includegraphics[width=0.7\linewidth]{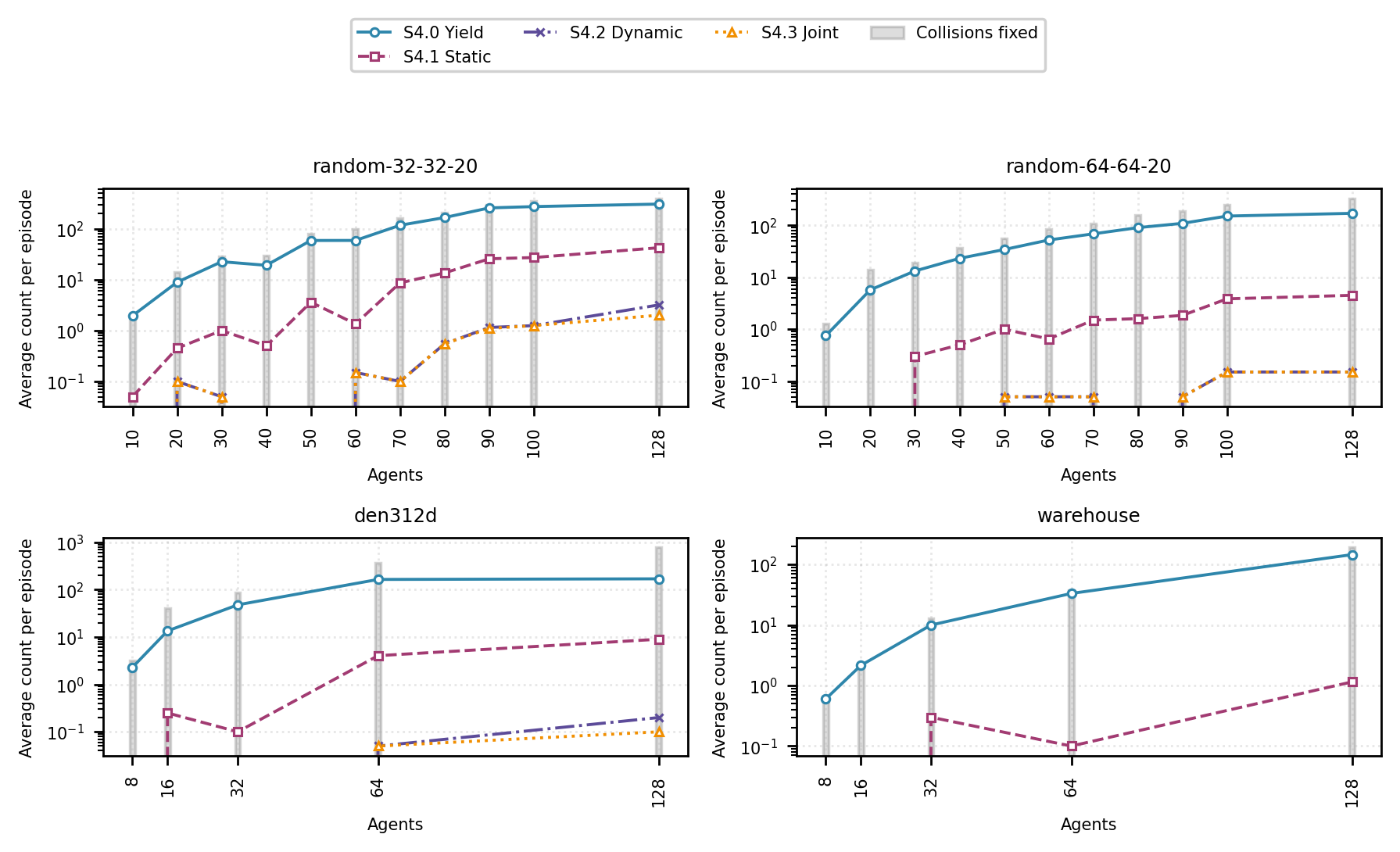}
    \caption{Strategy usage summary on random-32-32-20 (random-32), random-64-64-20 (random-64), den312d, and warehouse maps. Values are averaged across 20 problem instances.}
    \label{fig:strat_usage}
\end{figure*}

\begin{figure*}[!h]
    \centering
    \includegraphics[width=1\linewidth]{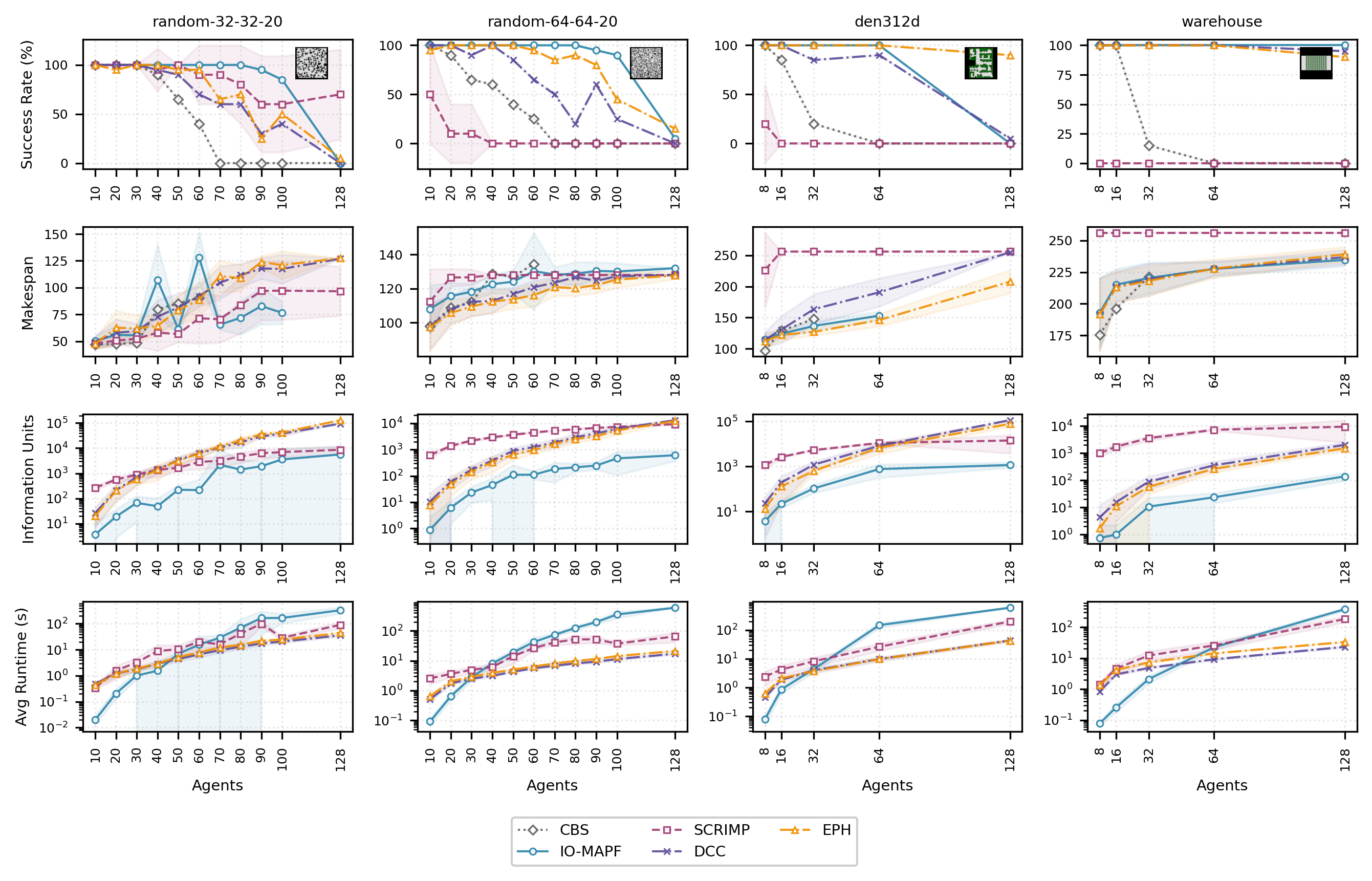}
    \caption{Performance of search-based (CBS, IC--MAPF) and learning-based (DCC, SCRIMP, EPH) MAPF solvers on random-32-32-20 (random-32), random-64-64-20 (random-64), den312d, and warehouse maps. Values are averaged across 20 problem instances. MS represents the makespan (maximum timesteps). Values are averaged within agent count groups. Lower MS$\downarrow$ and IU$\downarrow$, higher SR$\uparrow$ are better.}
    \label{fig:results_plot}
\end{figure*}

Tables~\ref{tab:search-vs-learning}, Figures~\ref{fig:strat_usage}, and Figure~\ref{fig:results_plot} provide expanded performance analyses complementing the compact results presented in the main paper. Table~\ref{tab:search-vs-learning} reports the full per-configuration statistics for all solvers across four benchmark maps, including success rate, makespan, and absolute values for Information Units (IU). These detailed values highlight the consistency of IC--MAPF across increasing agent densities and show the scale at which learning-based methods begin to degrade or saturate, especially in larger and more structured environments.

Figure~\ref{fig:strat_usage} summarizes the usage frequency of our tiered replanning strategies S4.0 to S4.3 (Yield, Static, Dynamic, and Joint) across all problem instances. This visualization illustrates how IC--MAPF automatically escalates strategy complexity only when needed, typically in high-density or tightly coupled regions, while relying predominantly on low-information repairs in easier configurations.

Figure~\ref{fig:results_plot} presents aggregate trends for \textit{success rate}, \textit{makespan}, \textit{IU consumption}, and \textit{average runtime}. The shaded regions indicate standard deviation across 20 problem instances, providing insight into solver stability and variance across different metrics. Compared to CBS and learning-based planners (DCC, SCRIMP, EPH), IC--MAPF achieves consistently high success rates while maintaining substantially lower information usage and competitive makespan behavior. Together, these additional results reinforce the main claims of the paper: IC--MAPF offers a favorable balance between scalability, robustness, and dramatically reduced inter-agent information requirements.

\begin{table*}[t]
  \centering
  \scriptsize
  \setlength{\tabcolsep}{3pt}
  \renewcommand{\arraystretch}{1.1}

  \begin{tabular}{c c c | *{14}{r}}
    \toprule
    & & & \multicolumn{2}{c}{Search-based} & \multicolumn{3}{c}{Hybrid} & \multicolumn{9}{c}{Learning-based Solvers} \\
    \cmidrule(lr){4-5} \cmidrule(lr){6-8} \cmidrule(lr){9-17}
    Map & & $m$ &
    \multicolumn{2}{c}{CBS} &
    \multicolumn{3}{c}{IC--MAPF} &
    \multicolumn{3}{c}{SCRIMP} &
    \multicolumn{3}{c}{DCC} &
    \multicolumn{3}{c}{EPH} \\
    & & &
    SR$\uparrow$ & MS$\downarrow$ &
    SR$\uparrow$ & MS$\downarrow$ & IU$\downarrow$ &
    SR$\uparrow$ & MS$\downarrow$ & IU$\downarrow$ &
    SR$\uparrow$ & MS$\downarrow$ & IU$\downarrow$ &
    SR$\uparrow$ & MS$\downarrow$ & IU$\downarrow$ \\
    \midrule

    \multirow{11}{*}{\includegraphics[height=1.5cm]{content/images/random-32-32-20}} &
    \multirow{11}{*}{\rotatebox{90}{\shortstack{\texttt{random}\\\texttt{32$\times$32-20}}}} &
    10 & \textbf{100} & \textbf{46.8} & \textbf{100} & 50.5 & \textbf{3.6} & \textbf{100} & 48.0 & 262.2 & \textbf{100} & 49.1 & 26.1 & \textbf{100} & \textbf{47.6} & 19.7 \\
     & &
    20 & \textbf{100} & \textbf{47.6} & \textbf{100} & 56.0 & \textbf{18.8} & \textbf{100} & 50.5 & 565.2 & \textbf{100} & 58.0 & 211.2 & 95 & 62.5 & 204.2 \\
     & &
    30 & \textbf{100} & \textbf{48.4} & \textbf{100} & 55.5 & \textbf{66.3} & \textbf{100} & 52.4 & 884.8 & \textbf{100} & 59.5 & 696.2 & \textbf{100} & 62.0 & 576.8 \\
     & &
    40 & 90 & 80.0 & \textbf{100} & 107.0 & \textbf{48.8} & 95 & \textbf{58.0} & 1377.7 & 95 & 72.8 & 1407.0 & \textbf{100} & 64.5 & 1403.0 \\
     & &
    50 & 65 & 85.3 & \textbf{100} & 61.4 & \textbf{221.4} & \textbf{100} & \textbf{56.8} & 1645.2 & 90 & 81.7 & 3356.1 & 95 & 79.0 & 3183.2 \\
     & &
    60 & 40 & 90.6 & \textbf{100} & 128.2 & \textbf{214.2} & 90 & \textbf{71.4} & 2822.8 & 70 & 92.0 & 6337.2 & 95 & 88.7 & 6155.1 \\
     & &
    70 & 0 & - & \textbf{100} & \textbf{65.5} & \textbf{2158.2} & 90 & 70.5 & 3158.8 & 60 & 104.5 & 10385.0 & 65 & 110.7 & 11735.3 \\
     & &
    80 & 0 & - & \textbf{100} & \textbf{71.8} & \textbf{1405.5} & 80 & 83.8 & 4660.8 & 60 & 111.7 & 17196.1 & 70 & 108.9 & 20330.8 \\
     & &
    90 & 0 & - & \textbf{95} & \textbf{82.8} & \textbf{1909.0} & 60 & 96.9 & 6292.6 & 30 & 117.8 & 29474.4 & 25 & 123.8 & 36040.6 \\
     & &
    100 & 0 & - & \textbf{85} & \textbf{76.5} & \textbf{3553.8} & 60 & 97.2 & 6982.9 & 40 & 117.5 & 37877.7 & 50 & 121.0 & 41383.7 \\
     & &
    128 & 0 & - & 0 & - & - & \textbf{70} & \textbf{96.6} & \textbf{8615.5} & 0 & 128.0 & - & 5 & 127.5 & 129548.7 \\
    \midrule

    \multirow{11}{*}{\includegraphics[height=1.5cm]{content/images/random-64-64-20}} &
    \multirow{11}{*}{\rotatebox{90}{\shortstack{\texttt{random}\\\texttt{64$\times$64-20}}}} &
    10 & 100 & 98.0 & \textbf{100} & 108.0 & \textbf{0.9} & 50 & 112.5 & 612.1 & \textbf{100} & 97.7 & 10.6 & 95 & \textbf{97.3} & 7.8 \\
     & &
    20 & 90 & 108.8 & \textbf{100} & 115.6 & \textbf{6.2} & 10 & 126.5 & 1382.0 & \textbf{100} & 107.8 & 59.4 & \textbf{100} & \textbf{105.7} & 47.8 \\
     & &
    30 & 65 & 112.0 & \textbf{100} & 118.2 & \textbf{23.4} & 10 & 126.7 & 2190.3 & 90 & 112.2 & 172.2 & \textbf{100} & \textbf{109.5} & 141.8 \\
     & &
    40 & 60 & 128.5 & \textbf{100} & 122.6 & \textbf{45.0} & 0 & 128.0 & - & \textbf{100} & 112.8 & 382.4 & \textbf{100} & \textbf{112.3} & 328.9 \\
     & &
    50 & 40 & 127.0 & \textbf{100} & 124.0 & \textbf{109.8} & 0 & 128.0 & - & 85 & 116.8 & 889.7 & \textbf{100} & \textbf{113.8} & 638.0 \\
     & &
    60 & 25 & 134.5 & \textbf{100} & 130.2 & \textbf{110.8} & 0 & 128.0 & - & 65 & 120.8 & 1258.0 & 95 & \textbf{116.1} & 964.6 \\
     & &
    70 & 0 & - & \textbf{100} & 128.2 & \textbf{183.0} & 0 & 128.0 & - & 50 & 123.5 & 1823.1 & 85 & \textbf{120.8} & 1640.2 \\
     & &
    80 & 0 & - & \textbf{100} & 128.7 & \textbf{211.9} & 0 & 128.0 & - & 20 & 126.8 & 2947.4 & 90 & \textbf{120.2} & 2520.7 \\
     & &
    90 & 0 & - & \textbf{95} & 130.4 & \textbf{245.4} & 0 & 128.0 & - & 60 & 124.6 & 4095.8 & 80 & \textbf{122.2} & 3187.4 \\
     & &
    100 & 0 & - & \textbf{90} & 130.1 & \textbf{462.1} & 0 & 128.0 & - & 25 & 127.2 & 6134.2 & 45 & \textbf{125.3} & 5290.0 \\
     & &
    128 & 0 & - & 5 & 132.0 & \textbf{607.6} & 0 & 128.0 & - & 0 & 128.0 & - & \textbf{15} & \textbf{127.7} & 12250.5 \\
    \midrule

    \multirow{5}{*}{\includegraphics[height=1.5cm]{content/images/den312d}} &
    \multirow{5}{*}{\rotatebox{90}{\shortstack{\texttt{den312d}\\\texttt{65$\times$81}}}} &
    8 & \textbf{100} & \textbf{97.2} & \textbf{100} & 115.3 & \textbf{3.6} & 20 & 226.2 & 1156.7 & \textbf{100} & 114.2 & 22.6 & \textbf{100} & 111.9 & 13.0 \\
     & &
    16 & 85 & 128.7 & \textbf{100} & 124.0 & \textbf{21.8} & 0 & 256.0 & - & \textbf{100} & 130.7 & 191.3 & \textbf{100} & \textbf{122.2} & 128.4 \\
     & &
    32 & 20 & 148.0 & \textbf{100} & 136.7 & \textbf{105.2} & 0 & 256.0 & - & 85 & 163.5 & 1195.6 & \textbf{100} & \textbf{127.0} & 627.5 \\
     & &
    64 & 0 & - & \textbf{100} & 153.1 & \textbf{759.9} & 0 & 256.0 & - & 90 & 190.7 & 8180.1 & \textbf{100} & \textbf{146.1} & 6722.3 \\
     & &
    128 & 0 & - & 0 & - & - & 0 & 256.0 & - & 5 & 254.8 & 112519.4 & \textbf{90} & \textbf{207.7} & \textbf{77108.9} \\
    \midrule

    \multirow{5}{*}{\includegraphics[height=1.5cm]{content/images/warehouse}} &
    \multirow{5}{*}{\rotatebox{90}{\shortstack{\texttt{warehouse}\\\texttt{161$\times$63}}}} &
    8 & \textbf{100} & \textbf{175.7} & \textbf{100} & 192.8 & \textbf{0.8} & 0 & 256.0 & - & \textbf{100} & 191.8 & 4.3 & \textbf{100} & 191.6 & 1.7 \\
     & &
    16 & \textbf{100} & \textbf{196.2} & \textbf{100} & 215.1 & \textbf{1.0} & 0 & 256.0 & - & \textbf{100} & 213.3 & 15.6 & \textbf{100} & 212.9 & 11.1 \\
     & &
    32 & 15 & 221.3 & \textbf{100} & 220.3 & \textbf{10.7} & 0 & 256.0 & - & \textbf{100} & 219.2 & 89.0 & \textbf{100} & \textbf{218.1} & 57.7 \\
     & &
    64 & 0 & - & \textbf{100} & \textbf{227.6} & \textbf{23.6} & 0 & 256.0 & - & \textbf{100} & 227.6 & 356.9 & \textbf{100} & 227.8 & 265.1 \\
     & &
    128 & 0 & - & \textbf{100} & \textbf{234.7} & \textbf{139.1} & 0 & 256.0 & - & 95 & 236.8 & 2016.5 & 90 & 239.1 & 1502.4 \\
    \bottomrule
  \end{tabular}
  \caption{Performance of search-based (CBS), hybrid (IC--MAPF), and learning-based (SCRIMP, DCC, EPH) MAPF solvers. MS represents makespan (maximum timesteps). Lower MS$\downarrow$ and IU$\downarrow$, higher SR$\uparrow$ are better.}
  \label{tab:search-vs-learning}
\end{table*}

\subsubsection{Conclusion}

This analysis highlights the difference in information architecture. The distributed method requires each of its 20 agents to continuously sense and process a heavy stream of local data, amounting to a total load of \textbf{10,260 IU}. In contrast, our hybrid method offloads this burden to a central coordinator, resulting in a total information load of only \textbf{691 IU}.

This represents a \textbf{$\sim$93\% reduction in the total information load}, quantifying the efficiency of our on-demand alert system. While our method has a central coordinator, the burden on each individual agent is drastically lower, as it does not require constant, high-bandwidth sensing of its environment.


\subsection{Properties of Tiered Replanning Strategies}
\label{apx:tiered-properties}

In this section, we summarize the soundness and (local) completeness properties of the tiered replanning operators introduced in S4 (Replanning). We assume a standard discrete-time MAPF model on a finite grid graph $G = (V,E)$ with a finite set of agents $\{a_1,\dots,a_N\}$. Time is discrete, $t = 0,1,\dots,T_{\max}$; each agent can either move along an edge $(u,v)\in E$ or wait at its current vertex. A \emph{plan} for agent $a_i$ is a finite sequence of vertices
\[
\pi_i = (v_i(0), v_i(1),\dots,v_i(T_i)),
\]
where $(v_i(t),v_i(t+1))$ is either an edge in $E$ or a wait, with $v_i(0)$ and $v_i(T_i)$ denoting its fixed start and goal positions, respectively. A \emph{joint plan} is the collection $\Pi = \{\pi_1,\dots,\pi_N\}$.

We adopt the standard MAPF collision model: a joint plan $\Pi$ is collision-free if it contains no vertex collision
\[
\exists i\neq j,\ \exists t:\ v_i(t) = v_j(t),
\]
and no edge collision
\[
\exists i\neq j,\ \exists t:\ v_i(t) = v_j(t+1)\ \wedge\ v_i(t+1) = v_j(t).
\]
Throughout this discussion, we assume that each tier uses an underlying search routine (single-agent planner or joint planner) that is complete on its induced finite state-time graph (e.g., BFS or A* with an admissible, consistent heuristic), and that any candidate repair is re-simulated against the other agents before being committed.

\paragraph{Soundness.}
We first formalize soundness of a replanning operator.

\begin{definition}[Soundness]
A replanning operator $\mathcal{R}$ is \emph{sound} if, whenever it returns a modified joint plan $\Pi'$, the resulting joint plan is collision-free under the given collision model.
\end{definition}

In our framework, S4.1--S4.4 operate under constraints provided by the control module (S3) and are only accepted if a global collision check succeeds.

\begin{lemma}[Soundness of S4.1--S4.4]
\label{lem:soundness}
Assume that each replanning operator (yield-based, bounded static, bounded dynamic, and local joint planning) only considers legal moves (graph edges or waits), obeys all constraints imposed by S3 (forbidden cells, reservations, and joint consistency), and that any candidate repaired plan is re-simulated against all other agents and discarded if any collision is detected. Then every committed repair produced by S4.1--S4.4 is collision-free; in particular, each tier is sound.
\end{lemma}

\begin{proof}[Proof sketch]
Each operator constructs candidate paths by exploring a constrained state-time graph in which illegal moves (violating static constraints, dynamic reservations, or internal joint consistency) are excluded. By construction, no candidate produced by the underlying planner violates these constraints. Before committing any repair, the resulting joint plan $\Pi'$ is simulated and checked for vertex and edge collisions against all other agents. If a collision is found, the candidate is rejected. Hence any repair that is actually committed has passed both constraint enforcement and a global collision check, and is therefore collision-free.
\end{proof}

\paragraph{Yield-based local coordination (S4.1).}
The yield operator chooses a single agent (possibly anchored at its start or goal) and constructs a short detour to a nearby ``parking'' cell, waits there while other agents traverse the conflicted region, and then rejoins the original plan. The search for a parking cell is restricted to a bounded local neighborhood and does not modify other agents' plans.

\begin{lemma}[Soundness and incompleteness of S4.1]
\label{lem:yield}
The yield-based operator S4.1 is sound but not complete even for resolving a single conflict: there exist MAPF instances in which a collision can be resolved by modifying agents' plans, but no admissible yield maneuver exists within the bounded neighborhood and single-agent restriction imposed by S4.1.
\end{lemma}

\begin{proof}[Proof sketch]
Soundness follows directly from Lemma~\ref{lem:soundness}: any candidate yield path is simulated and only committed if no collisions remain.

To see incompleteness, consider a long, narrow corridor of width one, and two agents at opposite ends in the middle of the corridor, wishing to swap positions. A known solution exists (e.g., coordinating their timing so one waits at the opening while the other passes), but under a strict yield model that only searches for a parking cell in a small neighborhood around the collision, it is impossible to find a nearby parking position that both satisfies the local constraints and removes the conflict. More generally, whenever resolving a conflict requires modifying the timing or paths of multiple agents together, or using parking locations outside the bounded neighborhood, S4.1 may fail to find a repair even though a global solution exists. Thus, S4.1 is sound but incomplete.
\end{proof}

\paragraph{Bounded static replanning (S4.2).}
In bounded static replanning, the control module S3 selects a single agent, fixes all other agents' plans, and designates a time window $\{t_{\mathrm{start}},t_{\mathrm{end}}\}$ along the agent's current plan. A static constraint set $\Delta_c \subseteq V$ encodes cells that must be avoided (e.g., conflict cells), and the planner recomputes only the affected suffix of the agent's trajectory while treating $\Delta_c$ as forbidden and all other agents' plans as fixed reservations.

This induced subproblem can be modeled as single-agent planning on a finite state-time graph whose nodes are $(v,t)$ and whose edges correspond to legal moves that respect static constraints and avoid collisions with fixed agents.

\begin{lemma}[Local completeness of S4.2]
\label{lem:static}
Consider the state-time graph induced for S4.2 over a finite horizon, with all vertices $(v,t)$ such that $v\in\Delta_c$ in the relevant window removed, and all nodes and edges that would collide with fixed agents pruned. Assume the single-agent planner used in S4.2 is complete on this graph (e.g., BFS or A* with an admissible, consistent heuristic). If there exists a collision-free path for the selected agent from its state at $t_{\mathrm{start}}$ to its target (goal or intermediate waypoint) within this graph, then S4.2 will find such a path and return a valid repair.
\end{lemma}

\begin{proof}[Proof sketch]
The state-time graph is finite, as both the underlying grid and the time horizon are finite. Every collision-free path that respects static constraints and avoids fixed agents corresponds to a path in this pruned graph. A complete search algorithm on a finite graph is guaranteed to find a path to the target node if one exists. Therefore, under the assumed planner, S4.2 is complete for its induced subproblem.
\end{proof}

\paragraph{Bounded dynamic replanning (S4.3).}
Bounded dynamic replanning extends S4.2 by treating portions of other agents' plans as time-varying obstacles (reservations) over a finite horizon around the conflict. In the corresponding state-time graph, nodes and edges that would cause vertex or edge collisions with these reservations are removed.

\begin{lemma}[Local completeness of S4.3]
\label{lem:dynamic}
Under the same assumptions as Lemma~\ref{lem:static}, but with additional time-indexed reservations derived from other agents' plans, bounded dynamic replanning (S4.3) is complete for its induced single-agent subproblem: if there exists a collision-free path that respects both static constraints and dynamic reservations in the finite horizon, the underlying complete planner will find it.
\end{lemma}

\begin{proof}[Proof sketch]
The state-time graph with dynamic reservations remains finite; reservations simply prune additional nodes and edges corresponding to disallowed states and transitions. Any path that respects all reservations corresponds to a path in this pruned graph. A complete search will discover such a path if it exists. Hence S4.3 is locally complete for its induced single-agent subproblem.
\end{proof}

\paragraph{Local joint planning (S4.4).}
When single-agent repairs are insufficient, S4.4 jointly replans a small subset of agents $S \subseteq \{1,\dots,N\}$ over a bounded time window around the conflict, treating all other agents as fixed reservations. The induced local problem can be modeled as a joint state-time graph whose nodes are $(\mathbf{v}(t), t)$, where $\mathbf{v}(t) = (v_k(t))_{k\in S}$ is a joint configuration with no internal vertex collisions and no collisions with reserved agents.

\begin{lemma}[Local completeness of S4.4]
\label{lem:joint}
Consider the joint state-time graph for a fixed subset of agents $S$ and a finite horizon $\{t_{\mathrm{start}},t_{\mathrm{end}}\}$, where nodes correspond to collision-free joint configurations of $S$ and edges correspond to legal joint moves (Cartesian products of individual moves) that do not introduce collisions among $S$ or with reserved agents. Assume that S4.4 runs a complete joint planner (e.g., joint A* with an admissible heuristic) on this graph. If there exists a collision-free joint path from the initial configuration of $S$ at $t_{\mathrm{start}}$ to the designated joint target region by $t_{\mathrm{end}}$, then S4.4 will find such a joint path.
\end{lemma}

\begin{proof}[Proof sketch]
The joint state-time graph is finite: the number of joint configurations is bounded by $|V|^{|S|}$ per time step, and the time horizon is finite. Edges are finite and defined by admissible joint actions. A complete search algorithm on this finite graph is guaranteed to find a path to a reachable goal region. Any valid joint plan corresponds to such a path. Therefore S4.4 is complete for its induced joint subproblem.
\end{proof}

\paragraph{Discussion and overall behavior.}
The above lemmas show that each tier is sound and that the static, dynamic, and joint operators (S4.2--S4.4) are complete for their respective induced subproblems under the assumption of complete search over the chosen finite horizon. Yield-based coordination (S4.1) is sound but intentionally incomplete. In the full algorithm, these tiers are combined with a deferral mechanism: if all tiers fail for an agent within the current attempt and bounded parameters, the agent is temporarily deferred (parked and excluded from further replanning), the remaining agents are resolved, and the deferred agents are reintroduced in a second phase with relaxed limits.

With fixed bounds on time windows, joint subset sizes, and search effort, the overall tiered strategy is best viewed as a sound, information-efficient repair framework with \emph{local} completeness guarantees, rather than a globally complete MAPF algorithm. In principle, completeness for the full MAPF instance could be recovered by allowing the temporal windows and sizes of jointly replanned subsets to grow as needed and by repeatedly applying the tiers (including deferral) until all conflicts are resolved; however, this configuration would come at a substantially higher computational cost and is not pursued in our current implementation.


\subsection{Hardware Results: TurtleBot4}
\label{app:robot_exp}

We evaluated IC--MAPF on five TurtleBot4 robots deployed in a $6 \times 6$ indoor grid. The centralized coordinator module (\textit{S3}) ran on a central host PC, while each robot also executed its own local planning (\textit{S1}) and replanning procedures (\textit{S4}) under isolated ROS~2 namespaces. Robots communicated only through a shared synchronization topic (\texttt{/sync\_barrier}) and operated under a ROS~2 Discovery Server setup, ensuring that no robot was configured to discover or access any other robot’s topics or services; the only cross-namespace visibility permitted was the synchronization channel. No onboard perception (LiDAR, RGB-D, or OAK-D) was used; instead, robots followed their assigned plans using open-loop odometry control. All remaining ROS interfaces, including odometry and velocity commands, remained fully namespace-isolated, ensuring minimal inter-agent information exchange.

At the start of each run, robots reset their poses via the \texttt{/reset\_pose} service and waited for odometry stabilization. Execution proceeded in strict lockstep: at every step, a robot (i) rotated to the required heading using closed-loop angular control, and (ii) moved forward by $0.45$\,m using odometry-based distance tracking. After completing its action, each robot published a synchronization message and waited until all robots reached the same step, ensuring coordinated, collision-free execution across the team. We evaluated five MAPF problems, each repeated five times, and logged communication statistics, execution times, and stepwise deviations from planned poses. Communication and execution metrics are reported in Table~\ref{app-tab:tb4_comm_summary_comm}, while execution-quality statistics are summarized in Table~\ref{app-tab:tb4_comm_summary_quality}.

These experiments demonstrate that IC--MAPF can be reliably deployed on real robots with extremely lightweight sensing and communication requirements. In contrast, learning-based MAPF methods that depend on field-of-view observations would require additional sensors, onboard neural inference, higher energy consumption, and greater bandwidth usage. IC--MAPF’s ability to operate using only minimal synchronization messages highlights its practicality for resource-constrained, scalable, and privacy-preserving multi-robot systems.

\begin{table}[t!]
    \centering
    \scriptsize
    \setlength{\tabcolsep}{5pt}
    \renewcommand{\arraystretch}{1.15}
    \begin{tabular}{c | c c c c}
        \toprule
        \textbf{Map} & \textbf{Plan.\ Msgs} & \textbf{IU} & \textbf{Exec.\ Msgs} & \textbf{Exec.\ Time (s)} \\
        \midrule
        Map~1 & 6  & 48 & 55 & 110.3 \\
        Map~2 & 16 & 15 & 75 & 162.6 \\
        Map~3 & 8  & 21  & 65 & 129.5 \\
        Map~4 & 8  & 14  & 65 & 145.7 \\
        Map~5 & 10  & 54 & 95 & 260.5 \\
        \bottomrule
    \end{tabular}
    \caption{Communication and execution-related statistics for five TurtleBot4 trials.}
    \label{app-tab:tb4_comm_summary_comm}
\end{table}

\begin{table}[t!]
    \centering
    \scriptsize
    \setlength{\tabcolsep}{6pt}
    \renewcommand{\arraystretch}{1.15}
    \begin{tabular}{c | c c c}
        \toprule
        \textbf{Map} & \textbf{MS} & \textbf{Pos.\ Err.\ (cm)} & \textbf{Head.\ Err.\ ($^\circ$)} \\
        \midrule
        Map~1 & 11 & 2.08 & 1.97 \\
        Map~2 & 15 & 2.63 & 2.42 \\
        Map~3 & 13 & 2.14 & 2.09 \\
        Map~4 & 13 & 2.12 & 2.01 \\
        Map~5 & 19 & 2.58 & 2.54 \\
        \bottomrule
    \end{tabular}
    \caption{Execution quality measures: makespan, average positional deviation, and heading deviation for five TurtleBot4 runs. Harder problems (higher MS) show slightly larger systematic drift.}
    \label{app-tab:tb4_comm_summary_quality}
\end{table}

\subsection{Literature review \label{ap: lit}}

\subsubsection*{Conflict-Based Search (CBS)}

\emph{S1 – Agent Planning: Decentralized}:
Each agent independently computes its path from the start to the goal using a single-agent pathfinding algorithm (e.g., $A^*$). This decentralized planning allows for efficient initial path computation without considering other agents.

\emph{S2 – Collision Detection: Centralized}:
After individual paths are planned, CBS centrally examines these paths to detect conflicts, such as two agents occupying the same location at the same time (vertex conflicts) or swapping positions simultaneously (edge conflicts). This centralized detection ensures systematic identification of all potential conflicts.

\emph{S3 – Collision Avoidance Policy: Centralized}:
Upon detecting a conflict, CBS resolves it by adding constraints to the agents' paths. Specifically, it creates two new branches in a constraint tree, each imposing a restriction on one of the conflicting agents to avoid the conflict. This centralized policy ensures optimal conflict resolution by exploring different constraint combinations \cite{Sharon2015}.

\emph{S4 – Agent Replanning: Decentralized}:
With new constraints in place, only the agents affected by these constraints replan their paths. Each affected agent independently computes a new path that adheres to the added constraints, maintaining the decentralized nature of the replanning process.


\subsubsection*{Large Neighborhood Search (LNS)}

\emph{S1 – Agent Planning: Centralized}:
LNS-based methods, such as MAPF-LNS2, begin with a centralized planning phase where initial paths for all agents are computed using a fast, suboptimal solver like Prioritized Planning (PP). This initial solution may contain conflicts but serves as a starting point for further refinement \cite{LiAAAI22}.

\emph{S2 – Collision Detection: Centralized}:
The system centrally identifies conflicts (e.g., vertex or edge collisions) in the initial set of paths. This global analysis ensures that all potential conflicts are detected and can be addressed in subsequent steps.

\emph{S3 – Collision Avoidance Policy: Centralized}:
LNS employs a centralized strategy to resolve conflicts by selecting subsets of agents (the "neighborhood") involved in collisions and replanning their paths. Techniques like Safe Interval Path Planning with Soft Constraints (SIPPS) are used to minimize the number of conflicts during this replanning phase.

\emph{S4 – Agent Replanning: Centralized}:
The selected subset of agents undergoes centralized replanning to resolve conflicts, while the paths of other agents remain unchanged. This process iterates, with different neighborhoods selected in each iteration, until a conflict-free set of paths is achieved or a predefined time limit is reached.


\subsubsection*{PRIMAL and PRIMAL2}

\emph{S1 – Agent Planning: Distributed}:
In both PRIMAL and PRIMAL2, each agent independently plans its path using policies learned through a combination of RL and IL. These policies are trained to enable agents to navigate towards their goals based on local observations without centralized coordination \cite{sartoretti2019primal, Damani2021PRIMAL2}.

\emph{S2 – Collision Detection: Distributed}:
Agents detect potential collisions based on their local observations of the environment. They do not rely on a centralized system to identify conflicts but instead use their learned policies to anticipate and respond to nearby agents.

\emph{S3 – Collision Avoidance Policy: Distributed}:
Collision avoidance is handled through the agents' learned behaviors. In PRIMAL2, enhancements such as improved observation types have been introduced to facilitate better implicit coordination among agents, especially in dense and structured environments.

\emph{S4 – Agent Replanning: Distributed}:
Agents continuously replan their paths in response to changes in their local environment. This reactive planning allows them to adapt to dynamic scenarios, such as new goal assignments in lifelong MAPF settings.


\subsubsection*{SCRIMP}

\emph{S1 – Agent Planning: Distributed}:
Each agent independently plans its path using a policy learned through a combination of RL and IL. Agents rely on a small local FOV (as small as 3×3) and a transformer-based communication mechanism to share information with nearby agents, enabling them to make informed decisions despite limited local observations \cite{Wang2023SCRIMP}.

\emph{S2 – Collision Detection: Distributed}:
Agents detect potential collisions based on their local observations and the messages received from neighboring agents through the communication mechanism. This decentralized approach allows agents to anticipate and respond to nearby agents without centralized coordination.

\emph{S3 – Collision Avoidance Policy: Distributed}:
Collision avoidance is handled through the agents' learned policies, which incorporate a state-value-based tie-breaking strategy. This strategy enables agents to resolve conflicts in symmetric situations by assigning priorities based on predicted long-term collective benefits and distances to goals.

\emph{S4 – Agent Replanning: Distributed}:
Agents continuously replan their paths in response to changes in their local environment, leveraging intrinsic rewards to encourage exploration and mitigate the long-term credit assignment problem. This decentralized replanning allows agents to adapt to dynamic scenarios effectively.


\subsubsection*{Learn to Follow (FOLLOWER)}

\emph{S1 – Agent Planning: Decentralized}:
Each agent independently plans its path to the assigned goal using a heuristic search algorithm (e.g., A*). To mitigate congestion, the planner incorporates a heatmap-based cost function that penalizes frequently occupied areas, encouraging agents to choose less crowded paths. A sub-goal (waypoint) is selected along the planned path to guide short-term movement \cite{Skrynnik2024}.

\emph{S2 – Collision Detection: Decentralized}:
Agents detect potential collisions based on their local observations. They do not rely on a centralized system to identify conflicts but instead use their learned policies to anticipate and respond to nearby agents.

\emph{S3 – Collision Avoidance Policy: Decentralized}:
Collision avoidance is handled through the agents' learned behaviors. A neural network-based policy guides the agent toward its sub-goal while avoiding collisions. The policy is trained using reinforcement learning, leveraging local observations without requiring global state information or inter-agent communication.

\emph{S4 – Agent Replanning: Decentralized}:
Agents continuously replan their paths in response to changes in their local environment. This reactive planning allows them to adapt to dynamic scenarios, such as new goal assignments in lifelong MAPF settings.


\subsubsection*{LNS2+RL}

\emph{S1 – Agent Planning: Centralized}:
LNS2+RL begins by centrally generating initial paths for all agents using a fast, suboptimal method like Prioritized Planning (PP). This initial solution may contain collisions but serves as a starting point for further refinement \cite{WangAAAI25}.

\emph{S2 – Collision Detection: Centralized}:
The system centrally identifies conflicts (e.g., vertex or edge collisions) in the initial set of paths. This global analysis ensures that all potential conflicts are detected and can be addressed in subsequent steps.

\emph{S3 – Collision Avoidance Policy: Hybrid}:
LNS2+RL employs a hybrid strategy for collision avoidance: \textbf{Early Iterations:} Utilizes a MARL policy to replan paths for subsets of agents involved in conflicts. This decentralized component allows agents to learn cooperative behaviors to avoid collisions. \textbf{Later Iterations:} Switches to a centralized PP algorithm for replanning, aiming to quickly resolve any remaining conflicts.

\emph{S4 – Agent Replanning: Hybrid}:
Replanning in LNS2+RL is conducted in a hybrid manner: \textbf{Early Iterations:} Selected subsets of agents undergo decentralized replanning using the MARL policy, promoting cooperative behavior. \textbf{Later Iterations:} Replanning shifts to a centralized approach using PP, focusing on efficiency and resolving any remaining conflicts.

\begin{table*}[h]
\scriptsize
\centering
\caption{Related Literature Categorization into -- S1: Agent Planning; S2: Collision Detection; S3: Collision avoidance policy; S4: Agent Replanning (SB: Search Based; LB: Learning Based). Our approach is IC--MAPF with S4 variants 4.0-4.3.}

\begin{tabular}{p{3cm} p{2.0 cm} p{2.0 cm} p{2.0cm} p{3.0cm}}
\toprule
\textbf{Method} & \textbf{S1} & \textbf{S2} & \textbf{S3} & \textbf{S4} \\
\toprule
\textbf{Our: IC--MAPF} & Decentralized & Centralized & Centralized & Distributed (4.0 - 4.3)\\
\midrule
(SB) CBS & Decentralized & Centralized & Centralized & Centralized \\
\midrule
(SB) LNS & Centralized & Centralized & Centralized & Centralized \\
\midrule
(LB) PRIMAL \& PRIMAL-2 & Distributed & Distributed & Distributed & Distributed \\
\midrule
(LB) SCRIMP & Distributed & Distributed & Distributed & Distributed \\
\midrule
(LB) Learn to Follow & Decentralized & Decentralized & Decentralized & Decentralized \\
\midrule
(LB) LNS2+RL & Centralized & Centralized & Hybrid & Hybrid \\
\bottomrule
\end{tabular}
\label{tab:lit_comparison}
\end{table*}

\subsection{Training Methods \label{ap: train}}
\subsubsection{PPO Training Procedure \label{ap: ppo}}

We train a single-agent navigation policy \(\pi_{\theta}\) to reach a specified goal in the presence of both static and dynamic obstacles. Dynamic obstacles are each assigned a hidden goal and follow a precomputed trajectory, executing only valid moves; this setup allows us to generate online collision alerts during inference (cf.\ stage S3 of our fix–collisions algorithm).

Training is performed with Proximal Policy Optimization (PPO) \cite{SchulmanWolskiDhariwalRadfordKlimov2017} on an \(11\times11\) maze grid for \(30{,}000\) episodes. Throughout the first \(15{,}000\) episodes, we linearly increase the static-obstacle density from \(0\%\) to \(30\%\) and the number of dynamic obstacles from 0 to 4.

At each time step \(t\), we compute the discounted return
\[
R_t = \sum_{l=0}^{T-t} \gamma^{l} \, r_{t+l},
\]
and the advantage estimate using Generalized Advantage Estimation (GAE) \cite{SchulmanMoritzLevineJordanAbbeel2015}:
\[
A_t = R_t - V_{\theta}(s_t),
\]
where \(\gamma=0.95\) and \(V_{\theta}(s_t)\) is the value function.

The PPO surrogate objective is

\begin{align}
L_{\mathrm{PPO}}(\theta) = -\mathbb{E}_t\big[\min\big( r_t A_t,\, \mathrm{clip}(r_t,1-\varepsilon,1+\varepsilon)\,A_t \big)\big]\nonumber\\
+ c_v\,\mathbb{E}_t\big[\big(V_{\theta}(s_t)-R_t\big)^2\big]
- c_e\,\mathbb{E}_t\big[H\big(\pi_{\theta}(\cdot\mid s_t)\big)\big]\nonumber,
\end{align}
where
$r_t = \frac{\pi_{\theta}(a_t\mid s_t)}{\pi_{\theta_{\mathrm{old}}}(a_t\mid s_t)}$,
\(\varepsilon=0.2\), \(c_v=0.5\), and \(c_e\) are linearly annealed from 0.05 down to 0.01 over the first \(5{,}000\) episodes. PPO inherits many of the stability guarantees of trust-region methods \cite{SchulmanLevineAbbeelJordanMoritz2015}.

To encourage the agent to distinguish valid from invalid moves, we add a binary cross-entropy loss
\[
L_{\mathrm{valid}}(\theta) = \mathbb{E}_t\big[ \mathrm{BCE}\big(z_t, m_t\big) \big],
\]
where \(z_t\in\mathbb{R}^{|\mathcal{A}|}\) are the network logits, \(m_t\in\{0,1\}^{|\mathcal{A}|}\) is the action-validity mask, and
\[
\mathrm{BCE}(z, m) = -\sum_{i=1}^{|\mathcal{A}|} \big[ m_i\log\sigma(z_i) + (1-m_i)\log\big(1-\sigma(z_i)\big) \big].
\]
The full objective is
\[
L(\theta) = L_{\mathrm{PPO}}(\theta) + \lambda_{\mathrm{valid}}\,L_{\mathrm{valid}}(\theta),
\]
where \(\lambda_{\mathrm{valid}}\) is chosen empirically. During both training and inference, we sample only actions flagged as valid by \(m_t\), which accelerates convergence and improves safety.

After training, the policy \(\pi_{\theta}\) is used in stage S1 to generate initial trajectories and in stage S4 to replan whenever the centralized collision detector (stage S2) issues an alert.


\subsubsection{Neural Network Architecture \label{ap: network}}
\subsubsection*{ResNetDQN Network Architecture for DDQN Training}
\label{ap:network_ddqn}

ResNetDQN is a residual‐network architecture for approximating the action‐value function \(Q^\pi(s,a)\) in grid‐based environments trained using the DDQN algorithm \cite{van2016deep}.  It combines a deep convolutional stem with residual blocks \cite{He2016CVPR}, early fusion of low‐dimensional features, hierarchical downsampling, and a late‐fusion MLP to output one Q‐value per action.

\textbf{Network Overview}
\begin{enumerate}
  \item \textbf{\textit{Input:}}  
    \((B,6,H,W)\) tensor comprising four binary masks (obstacle, agent, goal, dynamic) plus normalized \(x\)- and \(y\)-coordinate channels, and a \((B,3)\) low‐dim vector of \{\texttt{direction}, \texttt{distance}\}.
  \item \textbf{\textit{Conv Stem \& Residual Blocks:}}  
    \(3\times3\) conv (6→32 channels, stride=1, pad=1) + BatchNorm + ReLU,  
    then two \texttt{ResidualBlock(32→32)} with dilation rates 1 and 2.
  \item \textbf{\textit{Early Fusion:}}  
    Project low‐dim (3→16), tile to \((B,16,H,W)\), concat with conv features → 48 channels,  
    then \(3\times3\) conv (48→32) + BatchNorm + ReLU.
  \item \textbf{\textit{Downsampling Stage 1:}}  
    \(3\times3\) conv (32→64, stride=2, pad=1) + BatchNorm + ReLU,  
    followed by two \texttt{ResidualBlock(64→64)} with dilation rates 2 and 4.
  \item \textbf{\textit{Downsampling Stage 2:}}  
    \(3\times3\) conv (64→128, stride=2, pad=1) + BatchNorm + ReLU,  
    followed by two \texttt{ResidualBlock(128→128)} with dilation rates 4 and 1.
  \item \textbf{\textit{Global Pooling \& Late Fusion:}}  
    AdaptiveAvgPool2d → 128‐dim vector \(\mathbf{u}\).  In parallel, map low‐dim→256, map \(\mathbf{u}\to256\), concat →512 →256 via FC + ReLU.
  \item \textbf{\textit{Output:}}  
    Linear layer (256→\(|\mathcal{A}|\)) produces Q‐values.
  \item \textbf{\textit{Initialization:}}  
    All conv and linear weights: Xavier‐uniform; all biases: zero.
\end{enumerate}

\textbf{Implementation and Packages}  
\begin{itemize}
  \item \texttt{torch}, \texttt{torch.nn}, \texttt{torch.nn.functional}: define modules, layers, activations.
  \item \texttt{AdaptiveAvgPool2d}, \texttt{BatchNorm2d}, \texttt{Conv2d}, \texttt{Linear}: core building blocks.
\end{itemize}


\subsubsection*{ResNet-based Actor–Critic Architecture for PPO Training}
\label{ap:network_ppo}

The PPOActorCritic model shares the same ResNet encoder as ResNetDQN (conv stem, residual blocks, fusion, downsampling, pooling) \cite{He2016CVPR}, producing a 256‐dim embedding.  It splits into two GRU‐based heads for policy (actor) and value (critic), trained via PPO \cite{SchulmanWolskiDhariwalRadfordKlimov2017}.

\textbf{Network Overview}
\begin{enumerate}
  \item \textbf{\textit{Shared Encoder:}}  
    Follows Steps 1–4 from ResNetDQN, yielding a 256‐dim hidden \(\mathbf{h}_{\mathrm{shared}}\).
  \item \textbf{\textit{Actor Head:}}  
    GRUCell(256→256) updates hidden state \(\mathbf{h}^\pi_t\);  
    Linear(256→\(|\mathcal{A}|\)) produces action logits.
  \item \textbf{\textit{Critic Head:}}  
    GRUCell(256→256) updates hidden state \(\mathbf{h}^V_t\);  
    Linear(256→1) produces scalar state‐value estimate.
  \item \textbf{\textit{Initialization:}}  
    Same Xavier‐uniform for all weights, zero biases.
\end{enumerate}

\textbf{Implementation and Packages}
\begin{itemize}
  \item \texttt{torch}, \texttt{torch.nn}, \texttt{torch.nn.functional}: define encoder, GRUs, heads.
  \item \texttt{GRUCell}, \texttt{Linear}: recurrent and output modules.
\end{itemize}


\newpage
\subsubsection{Training Results \label{ap: train_res}}

\end{document}